\documentclass[12pt]{article}

\usepackage{amssymb} % symbols as greek letters
\usepackage{amsthm} % theorem environment and theorem style
\usepackage[english]{babel} % set language to english
\usepackage[utf8]{inputenc}
\usepackage{dsfont} % used for identity symbol = \mathds{1}
\usepackage{float}
\usepackage{hyperref} % enable hyperlinks
\usepackage{mathtools} % useful eg \coloneqq
\usepackage{mathrsfs} % mathscr
\usepackage{extarrows} %arrows
\usepackage{enumerate}   
\usepackage{mathabx}
\usepackage{color}
\numberwithin{equation}{section}

\newtheorem{theo}{Theorem}[section]
\newtheorem{lemma}[theo]{Lemma}

\newtheorem{prop}[theo]{Proposition}

\theoremstyle{definition}
\newtheorem{defi}[theo]{Definition}

\theoremstyle{remark}
\newtheorem*{rem}{Remark}

\newtheorem*{example}{Example}
\newcommand{\N}{\mathbb N}

\newcommand{\Z}{\mathbb Z}
\newcommand{\R}{\mathbb R}
\newcommand{\C}{\mathbb C}

\newcommand{\diff}{\,\mathrm{d}}
\newcommand{\eto}{\,\mathrm{e}}

\newcommand{\diag}{{\rm diag\,}}

\newcommand{\trace}{{\rm Tr\,}}
\newcommand{\im}{{\rm Im \,}}
\newcommand{\re}{{\rm Re \,}}

\newcommand{\sdet}{{\rm Sdet \,}}

\title{Supersymmetric Polar Coordinates \\
	with applications to the Lloyd model
}
\author{Margherita Disertori \footnote{Institute for Applied Mathematics
\& Hausdorff Center for Mathematics, 
University of Bonn, Endenicher Allee 60, 53115 Bonn, Germany
\newline
E-mails: disertori@iam.uni-bonn.de,  lager@iam.uni-bonn.de \newline
Funded by the Deutsche Forschungsgemeinschaft (DFG, German Research Foundation) - Projektnummer 211504053 -
SFB 1060.
}
 \ and \ 
Mareike Lager\footnotemark[2]}

\date{}

\begin{document}

\maketitle

\begin{abstract}
Spectral properties of random Schrödinger operators are encoded 
in the average of products of Greens functions.
For probability distributions with enough finite moments, 
the supersymmetric approach offers a useful dual representation.
Here we use supersymmetric  polar coordinates to derive a dual representation
that holds for general distributions.
We apply this result to study the density of states 
of the linearly correlated Lloyd model.
In the case of non-negative correlation,
we recover the well-known exact formula. 
In the case of linear  small negative interaction localized around one point,
we show that the density of states is well approximated by the exact formula.
Our results hold on the lattice $\Z^{d}$ uniformly in the volume. 
\end{abstract}

%Section 1: Introduction
\section{Introduction}

A major open problem in mathematical physics is the existence of an 
Anderson
transition in dimension three and higher for random Schrödinger operators.
These operators model transport in  disordered media, a
classical example being  electrical conductivity in metals with impurities.
In this paper, we consider the quantum mechanical problem of an electron
moving on a lattice $\Z^{d}$ and interacting with a random potential. 
The corresponding mathematical model is the so-called 
discrete Random Schrödinger operator, 
or Anderson's tight binding model \cite{anderson1958},
acting on the Hilbert space $l^{2} (\Z^{d})$ and defined by
\begin{align*}
H \coloneqq - \Delta_{\mathbb{Z}^{d}} + \lambda V,
\end{align*}
where $\Delta_{\mathbb{Z}^{d}}$ is the lattice Laplacian
$(\Delta \psi) (j)=\sum_{k:|j-k|=1} (\psi (j) -\psi (k))$,
and $V$ is a multiplication operator $(V\psi)(j) = V_j \psi(j)$.
Here, $\{V_{j} \}_{j\in \mathbb{Z}^d}$ is a collection of random variables 
(independent or correlated) 
and $\lambda>0$ is a parameter expressing the strength of disorder.
Physical information are encoded in the spectral properties of $H$.
For a large class of random potentials $V$ localization 
of the eigenfunctions has been proved
in $d=1$ for arbitrary disorder 
and in $d\geq 2$ for large disorder or at the band edge.
A localization - delocalization transition
has been proved on tree graphs, 
and is conjectured to hold on $\mathbb{Z}^{d},$ for $d\geq 3.$ 
A detailed up-to-date review on the model, known results and tools 
can be found in the book by Aizenman and Warzel \cite{AizWar}. 

Finite volume criteria allow to reconstruct properties of $H$ 
from the Green's function (or resolvent)
of a finite volume approximation $H_{\Lambda},$
by taking the thermodynamic limit $\Lambda \uparrow \mathbb{Z}^d.$
More precisely, let  $\Lambda\subset \Z^d$ be a finite cube
centered around the origin  with volume  $|\Lambda| =N$. 
We define the Random Schrödinger operator 
$H_{\Lambda}\in l^2(\Lambda)$ on $\Lambda$ as
\begin{align}\label{eq:Hfinitevol}
H_{\Lambda }= -\Delta+\lambda V ,
\end{align}
where $\Delta=\Delta_{\Lambda } $ is the discrete Laplacian on $\Lambda$
\begin{align*}
(\Delta \psi)(j)  = \sum_{k\in \Lambda :|j-k|=1} (\psi (k)- \psi (j))
+ \mbox{ eventual boundary terms. }
\end{align*}
The relevant quantities are expressions of the form 
\begin{align}\label{eq:prod}
\mathbb{E}[ G_{\Lambda}(z_{1})_{j_{1},k_{1}}\dots
G_{\Lambda}(z_{n})_{j_{n},k_{n}}],
\end{align}
where $G_{\Lambda } (z)\coloneqq (z\mathds{1}_{\Lambda }-H_{\Lambda })^{-1},$ 
$z\in \mathbb{C}\setminus \sigma (H),$ and 
$\mathbb{E}$ denotes the average with respect 
to the random vector $V.$

In particular the (averaged) density of states $\bar\rho_{\lambda } (E)$
satisfies the relation,
\begin{align*}
\int_{\R} \frac{1}{z-E}\ \bar\rho_{\lambda } (E)  \diff E
= \frac{1}{\pi|\Lambda|}\mathbb{E}[\trace G_{\Lambda }(z )],
\end{align*}
hence (see for example \cite[Section 4 and Appendix B]{AizWar})
\begin{align*}%\label{eq:model:averagedos}
\bar\rho_\Lambda(E)  
\coloneqq  -\frac{1}{\pi|\Lambda|} \lim_{\varepsilon\to 0^{+}} 
\mathbb{E}[ \im \trace G_{\Lambda }(E+i\varepsilon )],
\end{align*}
where $E\in\mathbb{R}.$ 
Regularity properties of $\bar\rho_\Lambda(E)$ and its derivatives 
can be inferred from the  generating function 
\begin{align}\label{eq:generatorG}
\mathcal{G}_{\varepsilon }(E,\tilde E ) =
\mathbb{E}\left[ 
\frac{\det ((E+ i\varepsilon) \mathds{1}_{\Lambda } - H_{\Lambda })}
{\det ((\tilde E + i \varepsilon)\mathds{1}_{\Lambda }- H_{\Lambda })}
\right].
\end{align}
For example 
\begin{align}\label{eq:gree-repr}
\trace G_{\Lambda }(E+i\varepsilon )
= -\partial_{\tilde E} \mathcal{G}_{\varepsilon }(E,\tilde E)|_{\tilde E= E}
=\partial_{ E} \mathcal{G}_{\varepsilon }(E,\tilde E)|_{\tilde E= E}.
\end{align}

Information on the nature of the spectrum can be deduced
from the thermodynamic limit of
\begin{align*}
\mathbb{E}[|G_{\Lambda }(E+i\varepsilon )_{jk} |^2],
\qquad \mbox{or}\qquad
\rho_2(E, E+\omega) 
\coloneqq \mathbb{E}[ \rho_\Lambda(E) \rho_\Lambda(E+\omega )]
\end{align*}
where the spectral parameter $\varepsilon $ and the energy difference $\omega $
must be taken of order $|\Lambda |^{-1}.$
\vspace{0,2cm}

A possible tool to analyse these objects is the so-called
supersymmetric (SUSY) approach. 
It allows to rewrite averages of the form \eqref{eq:prod}
as an integral involving only the Fourier transform
of the probability distribution, at the cost
of introducing Grassmann variables in the intermediate steps. 
A short introduction on Grassmann variables and their application
in our context is given in Appendix \ref{app:susy}.
For more details see for example the following monographs:
\cite{Varadarajan2004, Berezin1987,Wegner2016,DeWitt1992}.
This formalism proved to be especially useful in the case of random operators
arising from quantum diffusion problems \cite{Efetov1999}.
The supersymmetric approach was applied with success to study
Anderson localization  as well as  phase transitions on tree-graphs  
\cite{Wang2001,bovier1990,campanino1986supersymmetric,klein1986rigorous}.
All these applications are based on variations of the following key fact.

\begin{theo}\label{theo:generalrepresentation}
	Let $H_{\Lambda }$  be as in Eq. \eqref{eq:Hfinitevol} 
	and assume the $V_j$ are independent random variables with probability
	measure $\mu_{j}$ such that
	$\int v_{j}^{2}d\mu_{j} (v_{j})<\infty $ $\forall j$, i.e., 
	its Fourier transform 
	$\hat{\mu}_{j} (t)\coloneqq \int e^{-itv_{j}}d\mu_{j} (v_{j})$
	is twice differentiable with bounded first and second derivatives.
	
	Let $\mathcal{A} = \mathcal{A}[\{\chi_{j},\bar\chi_{j} \}_{j\in \Lambda}\}]$
	be a Grassmann algebra, $z\in \C^\Lambda$ a family of
	complex variables  and set 
	$\Phi_{j} \coloneqq (z_{j},\chi_{j})^t ,$  $\Phi^*_{j} \coloneqq (\bar z_{j}, \bar \chi_{j})$
	such that 
	$\Phi^*_{j}\Phi_{k}= \bar z_{j} z_{k}+ \bar \chi_{j}\chi_{k}$
	is an even element in $\mathcal{A}$ for all $j,k\in \Lambda $.
	For any matrix   $A\in \mathbb{C}^{\Lambda \times \Lambda }$, we define 
	\begin{align*}
	\Phi^* A\Phi
	\coloneqq \Phi^*  \diag (A, A)\Phi
	=\sum_{j,k\in \Lambda }A_{jk}\Phi^{*}_{j} \Phi_{k},
	\end{align*}
	where $ \diag (A, A)$ is a $2|\Lambda |\times 2 |\Lambda |$ 
	block diagonal matrix.
	In particular $\Phi^* \Phi= \sum_{j\in \Lambda }\Phi^{*}_{j} \Phi_{j}.$
	Finally, for any even element $a=b_{a}+n_{a}$ in $\mathcal{A}^0$
	with $n_{a}^{3}=0$ we define (cf. Eq. \eqref{eq:upgradefunc})
	\begin{align}\label{eq:fouriera}
	\hat{\mu}_j(a)
	= \mathbb{E}[e^{iaV_{j}}] 
	\coloneqq \hat{\mu}_j(b_{a})+\hat{\mu}'_j(b_{a}) n_{a}
	+\tfrac{1}{2}\hat{\mu}''_j(b_{a})n_{a}^{2} .
	\end{align}
	Then the generating function \eqref{eq:generatorG} can be written as
	\begin{align}\label{eq:susyav1}
	\mathcal{G}_{\varepsilon }(E,\tilde E) 
	= \int [\diff \Phi^* \diff \Phi]\ 
	\eto^{i \Phi^* (\mathbf{E} + i \varepsilon + \Delta)\Phi}
	\prod_{j\in \Lambda }\hat{\mu}_j(\lambda \Phi^*_j \Phi_j),
	\end{align}
	where we defined 
	$[\diff \Phi^* \diff \Phi] 
	\coloneqq \prod_{j\in \Lambda } (2\pi )^{-1} \diff \bar z_{j}  \diff z_{j}
	\diff \bar \chi_{j}  \diff \chi_{j}$, 
	$ \Phi^* \varepsilon \Phi=\varepsilon  \Phi^* \Phi$
	and $\mathbf{E} 
	= \diag (\tilde E \mathds{1}_{|\Lambda|}, E\mathds{1}_{|\Lambda|})$
	is a diagonal matrix. Moreover
	\begin{align}\label{eq:susyav2}
	\begin{split}
	\mathbb{E}[|G_{\Lambda }(E+i\epsilon )_{jk}|^2] 
	=&
	\int [\diff {\Phi}^* \diff \Phi]\, [\diff {\tilde \Phi}^* \diff \tilde \Phi]\, 
	\eto^{i {\Phi}^* (\mathbf{E} + i \varepsilon + \Delta)\Phi
		-i {\tilde \Phi}^* (\mathbf{E} -i \varepsilon + \Delta)\tilde \Phi} \\
	& \times z_{j}\bar z_k \tilde{z}_{k} \overline{\tilde z}_j\,
	\prod_{j\in\Lambda}\hat{\mu}_{j}
	(\lambda (\Phi_j^* \Phi_j- {\tilde \Phi_j}^* \tilde \Phi_j)).
	\end{split}
	\end{align}
	A similar representation holds for the two-point function $\rho_2(E,\tilde E)$.
\end{theo}

\begin{rem}
	In the formulas above both $\hat{\mu}_{j}(\lambda (\Phi_j^* \Phi_j))$ and
	$\hat{\mu}_{j}(\lambda (\Phi_j^* \Phi_j- {\tilde \Phi_j}^* \tilde \Phi_j))$
	are  well defined.
	Indeed, the even elements $a_{1}\coloneqq\Phi_j^* \Phi_j$ and
	$a_{2}\coloneqq\Phi_j^* \Phi_j- {\tilde \Phi_j}^* \tilde \Phi_j,$
	have nilpotent part
	$n_{a_{1}}=\bar \chi_{j}\chi_{j}$ and
	$n_{a_{2}}=\bar \chi_{j}\chi_{j}-\bar {\tilde{\chi}}_{j}\tilde{\chi}_{j} $,
	respectively.
	The result then follows from  $n_{a_{1}}^{2}=0=n_{a_{2}}^{3},$ 
	together with Eq. \eqref{eq:fouriera}.
	
	Note that we have taken independent variables above
	only to simplify notations.
	In the general case, the product of one-dimensional Fourier transforms 
	is replaced by a joint Fourier transform.
	The generalized formula will hold as long as the Fourier
	transform admits enough derivatives. 
\end{rem}

\begin{proof}
	We write $\mathcal{G}_{\varepsilon }(E,\tilde E) $ 
	and $\mathbb{E}[|G_{\Lambda }(E+i\varepsilon )_{jk}|^2]$
	as a supersymmetric integral (cf. Theorem \ref{theo:susyapproach})
	\begin{align*}
	&\mathcal{G}(E,\tilde E) 
	=  \mathbb{E}\left[ \int [\diff \Phi^* \diff\Phi]\,
	\eto^{i \Phi^* (\mathbf{E} + i \varepsilon + \Delta - \lambda V) \Phi}\right]\\
	&\mathbb{E}[|G_{\Lambda }(E+i\varepsilon )_{jk}|^2] =\\
	&\qquad \quad   \mathbb{E}\left[ \int [\diff \Phi^* \diff\Phi]\,
	\, [\diff {\tilde \Phi}^* \diff \tilde \Phi]\, 
	\eto^{i \Phi^* (\mathbf{E} + i \varepsilon + \Delta - \lambda V) \Phi
		-i {\tilde \Phi}^* (\mathbf{E} -i \varepsilon + \Delta-\lambda V)\tilde \Phi}
	z_{j}\bar z_k \tilde{z}_{k} \overline{\tilde z}_j\right ]
	\end{align*}
	This step holds for any choice of $V\in \R^{\Lambda}.$
	Note that we need two copies of SUSY variables to represent
	$\mathbb{E}[|G_{\Lambda }(E+i\varepsilon )_{jk}|^2]$.
	When $\diff \mu_{j}$ admits two finite moments, we can move the
	average inside. The result  follows.
\end{proof}

The aim of this paper is to extend this representation 
to probability distributions with less regularity.
To this purpose we introduce a supersymmetric version of polar coordinates
which allows to reexpress
$e^{i\lambda V_{j} \Phi^*_{j}\Phi_{j}}$ as $e^{i\lambda V_{j} x_{i}},$
where $x_{j}\in \R$ is a real variable.
As a result, the formula can be extended to any probability distribution
on $N=|\Lambda |$ real variables.
In contrast to the ordinary ones, 
supersymmetric polar coordinates introduce correction terms
due to the boundary of the integration domain.
The simple formula above will then be replaced by a sum of integrals.

As a concrete example, we consider the so-called Lloyd model,
with $V$ defined as  
$V_j\coloneqq \sum_{k\in \Lambda } T_{jk} W_k$, 
where $\{W_{k} \}_{k\in \Lambda }$
is a family of i.i.d. random variables with Cauchy distribution
$\diff \mu(x)= \pi^{-1}(1+x^2)^{-1} \diff x.$
The standard (uncorrelated) Lloyd model corresponds to $T_{jk}=\delta_{jk}.$ 
In this case the variables
$\{V_{j} \}_{j\in \Lambda }$ are independent and Cauchy distributed.
Note that $ \diff \mu (x)$ has no finite moments.
For this model, the averaged Green's function
(and hence the density of states) can be computed exactly whenever 
$T_{jk}\geq 0$ $\forall j,k$  (non-negative correlation)
\cite{Lloyd1969,Simon1983}.

Using supersymmetric polar coordinates, we show here that
for the non-negative linearly correlated Lloyd model Eq. 
\eqref{eq:susyav1} and  \eqref{eq:susyav2} remain valid,
with an appropriate redefinition of $\hat{\mu} (b_{a}+n_{a}).$
In this case, one can easily recover the exact formula 
for the averaged Green's function.
The formula remains valid also in the case of linear negative correlation,
at the price of adding  additional correction
terms, due to boundary effects.

We expect the supersymmetric representation will help to study 
problems not yet accessible via other tools, such as negative correlations
or the two point function at weak disorder.
As a first test, we considered  a simplified model
with small negative correlations localized on one site.
For this toymodel we used the supersymmetric representation to prove
that the density of states remains in the vicinity of the
exact formula. Our result holds in any dimension and arbitrary volume.

\paragraph{Overview of this article.}
In Section \ref{sec:main} we state the main results of the paper,
and give some ideas about the proofs.
More precisely, Section  \ref{sec:main1} introduces 
supersymmetric polar coordinates
(Theorem \ref{theo:polar}),  with a general integrated function $f,$
not  necessarily  compactly supported. 
Applications to  $\mathcal{G}_{\varepsilon }(E,\tilde E) $ and
$\mathbb{E}[|G_{\Lambda }(E+i\varepsilon )_{jk}|^2]$ 
are given in Theorem \ref{theo:generalpolar}.
The detailed proofs of both theorems can be found in Section \ref{sec:polar}.
In Subsection \ref{sec:main2} we consider the Lloyd model 
and give an application of the formula for a simple toymodel.
The corresponding proofs are in Section \ref{sec:application}.

%Section 2: Main Results
\section{Main results} \label{sec:main}

\subsection{Supersymmetric polar coordinates}\label{sec:main1}

For an introduction to the supersymmetric formalism 
see Appendix \ref{app:susy}.

Consider first $\mathcal{A} [\bar \chi,\chi]$ 
a Grassmann algebra with two generators. 
The idea of supersymmetric polar coordinates is to transform between generators
$(\bar z, z,\bar\chi,\chi)$ of 
$\mathcal{A}_{2,2}(\C)$ and $(r,\theta,\bar \rho,\rho)$ of
$\mathcal{A}_{2,2}(\R^+\times(0,2\pi))$
\footnote{cf. Definition \ref{def:generators}.
	Note that $\bar \rho,\rho \in \mathcal{A}^1[\bar \chi,\chi]$.} 
such that $\bar z z + \bar \chi \chi = r^2$.
A reasonable change is
\begin{align}\label{eq:polarcoordinates}
\Psi  (r,\theta,\bar\rho,\rho)=  
\begin{pmatrix}
z (r,\theta,\bar\rho,\rho)\\
\bar z (r,\theta,\bar\rho,\rho)\\
\chi(r,\theta,\bar\rho,\rho) \\
\bar\chi(r,\theta,\bar\rho,\rho) 
\end{pmatrix}
\coloneqq
\begin{pmatrix}
\eto^{i\theta } (r-\tfrac12 \bar\rho \rho )\\
\eto^{-i\theta } (r-\tfrac12 \bar\rho \rho )\\
\sqrt{r}  \rho \\
\sqrt{r} \bar\rho 
\end{pmatrix}
\end{align}
Indeed, we have 
$\bar z z + \bar \chi \chi  
= (r-\frac 12 \bar\rho \rho)^2 + r \bar\rho\rho=r^2$.

Note that $0$ is a boundary point for polar coordinates 
since it maps $\R^+\times (0,2\pi)$ to $\C\backslash \{0\} $. 
For functions with compact support in $U = \C \backslash \{0\}$
a SUSY version of the standard coordinate change formula applies, 
where the Jacobian is replaced by a Berezinian,
c.f. Theorem \ref{theo:changeofvariables}.
On the contrary, functions with $f(0) \neq 0$ have no compact support in the domain 
$U= \C\backslash \{0\}$ and we collect additional boundary terms 
as the following theorem shows.

\begin{theo}[Supersymmetric polar coordinates]\label{theo:polar}
	Let $N\in\N$, $\mathcal{A}_{2N}$ the complex Grassmann algebra 
	generated by $\{\bar\chi_j ,\chi_j \}_{j=1}^N$ and 
	$\{\Phi_j^*,\Phi_j\}_{j=1}^N$ a set of supervectors 
	defined as in Theorem \ref{theo:generalrepresentation}.
	Let $f\in \mathcal{A}_{2N,2N}(\C^N)$ be integrable,
	i.e., all $f_I:\C^N\to \C$ are integrable.
	Then
	\begin{align}\label{eq:Ipolarsum}
	I(f) = \int_{\C^N} [\diff  \Phi^* \diff \Phi] \  f (\Phi^*,\Phi) =
	\sum_{\alpha\in\{0,1\}^N} I_\alpha(f)
	\end{align}
	with multiindex $\alpha$ and
	\begin{align}\label{eq:I2polar}
	I_\alpha(f) =  \pi^{-|1-\alpha|}
	\int_{(\R^+\times(0,2\pi))^{1-\alpha}}
	(\diff r \diff \theta \diff \bar \rho \diff \rho)^{1-\alpha}
	\ f\circ \Psi_\alpha (r,\theta, \bar\rho, \rho),
	\end{align}
	where  
	$(\diff r)^{1-\alpha}= \prod_{j:\alpha_j=0}\diff r_j$ and $\Psi_\alpha$
	is given by 
	$\Psi_\alpha : (r,\theta,\bar\rho,\rho) \mapsto (z,\bar z,\chi,\bar \chi)$ with
	\begin{align*}
	\begin{cases}
	z_j(r_j,\theta_j,\bar\rho_j,\rho_j)
	&=\delta_{\alpha_j0 }\, \eto^{i\theta_j} (r_j-\tfrac12 \bar\rho_j\rho_j),\\
	\bar z_j(r_j,\theta_j,\bar\rho_j,\rho_j)
	&=\delta_{\alpha_j0 }\, \eto^{-i\theta_j} (r_j-\tfrac12 \bar\rho_j\rho_j),\\
	\chi_j (r_j,\theta_j,\bar\rho_j,\rho_j)
	&= \delta_{\alpha_j0 }\, \sqrt{r_j} \rho_j,\\
	\bar\chi_j (r_j,\theta_j,\bar\rho_j,\rho_j)
	&= \delta_{\alpha_j0}\, \sqrt{r_j} \bar\rho_j.
	\end{cases}
	\end{align*}
\end{theo}

\begin{proof}
	See Section \ref{sec:polar}.
\end{proof}

\begin{rem}
	For $f$ compactly supported on $\C\backslash \{0\}$ 
	(this means in particular $f(0) = 0$), 
	we recover the result of Theorem \ref{theo:changeofvariables}.
	Namely for $\alpha = 0$, we obtain the right-hand side 
	of Theorem \ref{theo:changeofvariables}
	while all contributions from  $\alpha \neq 0$ vanish.
\end{rem}

\begin{example} 
	To illustrate the idea behind the above result, 
	consider the following simple example.
	Let $\varphi$ be the smooth compactly supported function 
	$\varphi:\R\to\R$, given by 
	\begin{align*}
	\varphi(x) = 
	\begin{cases}
	\eto^{-(1-2|x|^2)^{-1}} & \text{if }|x|<\frac{1}{\sqrt{2}}\\
	0 & \text{otherwise}.
	\end{cases}
	\end{align*}
	Note that $\varphi(0) = \eto^{-1} \neq 0,$ 
	hence $f(\bar z,z,\bar \chi,\chi) = \varphi(\bar z z + \bar \chi \chi)$ 
	is a smooth function without compact support in $\C\backslash\{0\}$.
	By a straightforward computation, we have
	\begin{align*}
	I(f) &= \int_{|z|<\frac{1}{\sqrt 2}} 
	[ \diff \Phi^* \diff \Phi]  \ 
	\eto^{-(1-2\bar z z)^{-1}} (1-2 (1-2\bar z z)^{-2} \bar\chi \chi )\\
	&= \frac{1}{2\pi}\int_0^{\frac{1}{\sqrt 2}} 
	\diff r \int_0^{2\pi} \diff \theta \ 
	4r \eto^{-(1-2r^2)^{-1}}  (1-2r^2)^{-2} =  e^{-1},
	\end{align*}
	where we expand the expression in the Grassmann variables 
	and change to ordinary polar coordinates after integrating
	over the Grassmann variables.
	Applying formulas \eqref{eq:Ipolarsum} and \eqref{eq:I2polar},
	we obtain directly
	\begin{align*}
	I (f)
	=  \pi^{-1}\int_{\R^+\times(0,2\pi)}  \mkern-30mu 
	\diff r \diff \theta \diff \bar \rho \diff \rho \ 
	f\circ \Psi (r,\theta,\bar\rho,\rho) +  \ f \circ \Psi (0)
	=  e^{-1},
	\end{align*}
	where the first integral vanishes, since $f\circ \Psi$ 
	is independent of $\bar\rho$ and $\rho$.
\end{example}

Now consider the generating function \eqref{eq:generatorG}.
In the case of an integrable density without other regularity conditions, 
we obtain  the following result.

\begin{theo}\label{theo:generalpolar}
	Let $\Lambda\subset \Z^{d}$ be a finite volume and 
	$H_{\Lambda } = -\Delta + \lambda V$
	be the Schrödinger operator introduced in Eq. \eqref{eq:Hfinitevol}, 
	where  $\{V_{j} \}_{j\in \Lambda }$ is a family of real random variables 
	with integrable joint density $\mu$. 
	Then the generating function \eqref{eq:generatorG} can be written as
	\begin{align}\label{eq:polarrepresentation}
	\mathcal{G}_\varepsilon(E,\tilde E) 	
	=\hspace{-0,4cm} \sum_{\alpha\in\{0,1\}^\Lambda}  
	\int_{(\R^+\times(0,2\pi))^{1-\alpha}} \mkern-40mu
	(\tfrac{\diff r \diff \theta \diff \bar \rho \diff \rho}{\pi})^{1-\alpha}
	\hat{ \mu}(\{\lambda r_j^2\}_{j\in\Lambda})|_{r^\alpha = 0}\,
	g\circ \Psi_\alpha (r,\theta,\bar \rho,\rho )
	\end{align}
	where $g (\Phi^*,\Phi) 
	= \eto^{i \Phi^* (\mathbf{E} + i \varepsilon + \Delta)\Phi}$,  
	$\mathbf {E} 
	= \diag (\tilde E \mathds{1}_{|\Lambda|}, E\mathds{1}_{|\Lambda|})$ 
	and $\hat{ \mu}(\{\lambda r_j^2\}_{j\in\Lambda})$ 
	is the $|\Lambda|$-dimensional, joint Fourier transform of $\mu$.
	Similarly
	\begin{align*}
	&\mathbb{E}[|G_\Lambda(E+ i\varepsilon)_{jk}|^2]\\
	&=
	\sum_{\substack{\alpha\in\{0,1\}^\Lambda\\
			\tilde\alpha\in\{0,1\}^\Lambda}}\pi^{-|1-\alpha| - |1-\tilde \alpha|}
	\int_{\substack{(\R^+\times(0,2\pi))^{1-\alpha}\\
			\times (\R^+\times(0,2\pi))^{1-\tilde\alpha}}}
	(\diff r \diff \theta \diff \bar \rho \diff \rho)^{1-\alpha}
	(\diff \tilde r  \diff \tilde \theta
	\diff \bar{\tilde \rho} \diff \tilde\rho)^{1-\tilde\alpha}\\
	&\qquad 
	\hat{ \mu}(\{\lambda (r_j^2-\tilde r_j^2)\}_{j\in\Lambda})_{r^\alpha=0 = \tilde r^{\tilde\alpha}}\,
	g^+\circ \Psi_\alpha (r,\theta,\bar \rho,\rho ) \ 
	g^-\circ \Psi_{\tilde\alpha}(r,\theta,\bar \rho,\rho ),
	\end{align*}
	where $g^+ (\Phi^*,\Phi) 
	= \bar z_k z_j\eto^{i \Phi^* (\mathbf{E} + i \varepsilon + \Delta)\Phi}$ and 
	$g^- (\tilde\Phi^*,\tilde\Phi) = 
	\overline{\tilde z_j} \tilde z_k
	\eto^{-i {\tilde\Phi}^* (\mathbf{E} - i \varepsilon + \Delta)\tilde\Phi}$.
\end{theo}

\begin{proof}[Idea of the proof]
	Again we write $\mathcal{G}_\varepsilon(E,\tilde E) $ and 
	$|G_\Lambda(E+i\varepsilon)_{jk}|^2$ as a supersymmetric integral 
	(Theorem \ref{theo:susyapproach}). 
	Note that we need two copies of SUSY variables to represent
	$|G_\Lambda(E+i\varepsilon)_{jk}|^2$. 
	Taking the average inside at this point would cause problems. 
	Hence we apply first our polar-coordinate formula Theorem \ref{theo:polar}.
	Since $r$ is now real, the expression 
	$\mathbb{E}[\eto^{i\lambda \sum_{j} V_j r_j^2}] $ is the
	standard Fourier transform  $\hat{ \mu}(\{\lambda r_j^2\}_{j\in\Lambda})$.
	Details can be found in Section \ref{sec:polar}.
\end{proof}

\subsection{Applications  to the Lloyd model}\label{sec:main2}
As a concrete example, we consider the Lloyd model with linear correlated random potentials, i.e.
$V_j= \sum_{k} T_{jk} W_k$, where
$W_k \sim \text{Cauchy}(0,1)$ are i.i.d. random variables, $T_{jk}=T_{kj} \in \R$ and $\sum_{j} T_{jk}> 0$.

We discuss three cases:
\begin{enumerate}
	\item the classical Lloyd model, where $T_{jk} = \delta_{jk}$, hence $V_j \sim \text{Cauchy}(0,1)$ are i.i.d.
	\item the (positive) correlated Lloyd model, where  $T_{jk} \geq 0$ with $\sum_{j}T_{jk} >0$.
	\item a toymodel with single  negative correlation, i.e.  $T_{jj} = 1$ and $T_{21} = T_{12} = -\delta^2 $ with $0<\delta < 1$ and $T_{jk} = 0$ otherwise. The indices $1$ and $2$ denote two fixed, nearest neighbour points $i_{1},i_{2}\in \Lambda$ with $|i_{1}-i_{2}|=1.$
\end{enumerate}

\begin{prop}\label{prop:Cauchyrepresentations} When  $T_{jk} \geq 0 $ for all $j,k$ (Case 1. and 2. above) we have 
	\begin{align*}%\label{eq:prop}
	\mathcal{G}_\varepsilon (E,\tilde E) 	
	= \int [\diff \Phi^* \diff \Phi ]\ 
	g(\Phi^*,\Phi)
	\eto^{-\sum_{k}\lambda \sum_{j} T_{jk}\Phi_j^* \Phi_j}. 
	\end{align*}
	where $g(\Phi^*,\Phi) \coloneqq\eto^{i \Phi^* (\mathbf{ E} - i \varepsilon + \Delta)\Phi}$.
	For the toymodel (Case 3.  above) a similar formula holds with additional correction  terms. Precisely 
	\begin{align*}%\label{eq:proptoy}
	\begin{split}
	\mathcal{G}_\varepsilon (E,\tilde E) 	
	= \hspace{-0,2cm}&\sum_{ \beta \in \{++,+-,-+\} }  \int_{{ \mathcal{I}}_\beta}[\diff \Phi^* \diff \Phi] \ h(\Phi^*,\Phi) \eto^{-\lambda \sum_{j=1}^2  T^\beta_j\Phi_j^* \Phi_j} + R(h)
	\end{split}
	\end{align*}
	where $h (\Phi^*, \Phi) = g(\Phi^*,\Phi) \eto^{-\lambda \sum_{j\neq 1,2}\Phi^*_j \Phi_j}$, we defined	$T^{++} = (1-\delta^2)(1,1) $,
	$T^{+-} = (1+\delta^2)(1,-1)$ and 
	$T^{-+} = (1+\delta^2)(-1,1)$ and  \begin{align}\label{eq:Ibeta}
	\begin{split}
	\mathcal{I}_{++} &=
	\{z\in \C^N : \delta |z_2| < |z_1| <  |z_2|/\delta\} ,\\
	\mathcal{I}_{+-} &= 
	\{z\in \C^N :  |z_1| >  |z_2|/\delta\} ,\\
	\mathcal{I}_{-+} &= 
	\{z\in \C^N : |z_1| < \delta |z_2|\} .
	\end{split}
	\end{align}
	Moreover, the additional boundary term is given by
	\begin{align}\label{eq:R(h)}
	\begin{split}
	R(h)= & -\frac{1}{\pi^2}\int_{\R^+ \times (0,2\pi)^2} \diff r_2 \diff \theta_1 \diff \theta_2 [\diff \hat{\Phi}^* \diff \hat \Phi] \ h \circ \Psi_{12}(r_2,\theta_1,\theta_2, \hat\Phi^*,\hat\Phi) \\
	&\times \lambda r_2 \delta^2 \left[\eto^{-\lambda (1-\delta^4)r_2^2} +\delta^2 \eto^{-\lambda (1-\delta^4)\delta^2 r_2^2} \right],
	\end{split}
	\end{align}
	where $\hat\Phi = (\Phi_j)_{j\in \Lambda \backslash \{1,2\}}$ and 
	\begin{align*}
	\Psi_{12} (r_2,\theta_1,\theta_2) 
	= (\Phi_1, \Phi_2)
	= 
	\begin{pmatrix}
	\eto^{i \theta_1} \delta r_2 & \eto^{i \theta_2} r_2 \\ 0& 0\\
	\end{pmatrix} 
	\end{align*}
	The same formulas hold for $\mathbb{E}[\trace G_\Lambda(E+i \varepsilon)]$ with $g$ replaced by $g_1 (\Phi^*,\Phi) =\sum_{j} |z_j|^2 \eto^{i \Phi^* ( E - i \varepsilon + \Delta)\Phi}.$
\end{prop}

\begin{proof}[Idea of the proof]
	We  use the representation from Theorem \ref{theo:generalpolar} and  insert the Fourier transform of the given
	density.  For non-negative correlations we can then undo the coordinate change. When negative correlations are present this operation generates additional correction terms.  For details see Section~\ref{sec:application}.
\end{proof}

\noindent In the case of non-negative correlations we recover exact formulas, as follows. 
\begin{theo}\label{theo:mainresult}
	Let $T_{jk}=\delta_{jk}$ (classical Lloyd model). We have
	\begin{align}\label{eq:mainresult}
	\lim_{\varepsilon\to 0}\mathcal{G}_\varepsilon(E,\tilde E) = \frac{\det ((E+ i\lambda)\mathds{1}_\Lambda - H_0)}{\det ((\tilde E+ i\lambda)\mathds{1}_\Lambda - H_0)},
	\end{align}
	where $H_0 = -\Delta$. In particular 
	\begin{align}\label{eq:mainresult2}
	\lim_{\varepsilon\to 0}\mathbb{E}[ \trace G_\Lambda(E+ i\varepsilon)] =  \trace ((E+ i\lambda)\mathds{1}_\Lambda - H_0)^{-1}.
	\end{align}
	For $T_{jk}\geq 0$ (non-negative  correlation)  Eq. \eqref{eq:mainresult} and \eqref{eq:mainresult2} still hold, with
	$\lambda\mathds{1}_\Lambda$  replaced
	by the diagonal matrix $\lambda \hat T$, where $\hat T_{ij} = \delta_{ij}  \sum_{k} T_{jk}$.
	
	In particular both, the classical and the (positive) correlated Lloyd model have the same (averaged) density of states as the
	free Laplacian $H_0=-\Delta$ with imaginary mass $\lambda$ and $\lambda \hat T$, respectively.
\end{theo}

\begin{proof}[Idea of the proof]
	Follows from Proposition \ref{prop:Cauchyrepresentations}. For details see Section \ref{sec:application}.
\end{proof}

Note that the results on the density of states above can be derived also by other methods (cf. \cite{Lloyd1969} and \cite{Simon1983}).

In the case of localized negative correlation (the toymodel in Case 3. above) we obtain the following result.

\begin{theo}[Toymodel]
	\label{theo:toymodel}
	Consider $T_{jk}$ be as in Case 3. above, $\lambda >0$ and 
	$0<\delta \ll (1+\lambda^{-1})^{-1}.$ Then 
	\begin{align*}
	&\lim_{\varepsilon\to 0}\mathbb{E}[ \trace G_\Lambda(E+ i\varepsilon)] =\\
	&\mkern100mu 	\trace (E\mathds{1}_\Lambda +i \lambda \hat T- H_0)^{-1} \left [1+
	\mathcal{O}\Big ((\delta(1+\lambda^{-1}))^{2}\Big )+\mathcal{O} (|\Lambda |^{-1})\right ].\nonumber
	\end{align*}
\end{theo}

\begin{proof}[Idea of the proof]
	Follows from Proposition \ref{prop:Cauchyrepresentations} by integrating first over the uncorrelated variables in $\Lambda$ and estimating the remaining integral. For details see Section \ref{sec:application}.
\end{proof}

%Section 3: Supersymmetric polar coordinates}
\section{Supersymmetric polar coordinates}\label{sec:polar}

\subsection{Proof of Theorem \ref{theo:polar}}

\begin{proof}[Proof of Theorem \ref{theo:polar}]
	The idea is to apply the coordinate change $\Psi$ from Eq. \eqref{eq:polarcoordinates} for each $j\in\{0,\dots,N\}$.
	To simplify the procedure, we divide it into $\Psi_1\circ\Psi_2\circ\Psi_3$,
	where $\Psi_1$ is a change from ordinary polar coordinates into complex variables,  
	$\Psi_2$ rescales the odd variables
	and  $\Psi_3$ translates the radii into super space.
	Note that only the last step mixes ordinary and Grassmann variables and produces boundary terms.
	
	We first change the complex variables $z_j,\bar z_j$ for all $j$ into polar coordinates
	\begin{align*}
	\psi_1:(0,\infty) \times [0,2\pi) &\to \C\backslash \{0\}\\
	(r,\theta)& \mapsto z(r,\theta), \quad z_j(r_j,\theta_j) = r_j \eto^{i\theta_j} \quad\forall j.
	\end{align*}
	The Jacobian is $\prod_{j=1}^N 2r_j$ and by an ordinary change of variables
	\begin{align*}
	I(f)= \frac{1}{(2\pi)^N} \int_{(\R^+\times(0,2\pi))^N} \diff r \diff \theta \diff \bar\chi \diff \chi
	\prod_{j=1}^N 2r_j \ f \circ \Psi_1 (r,\theta,\bar \chi, \chi),
	\end{align*}
	where $\Psi_1 = \psi_1 \times \mathds{1}$.
	Note that no boundary terms arise.
	Now we rescale the odd variables by 
	\begin{align*}
	\psi_2(\bar\rho,\rho) \coloneqq(\bar\chi(\bar\rho,\rho),\chi(\bar\rho,\rho)) \quad 
	\begin{cases} \bar\chi_j (\bar\rho_j,\rho_j) \coloneqq  \sqrt{r_j} \bar\rho_j \\
	\chi_j  (\bar\rho_j,\rho_j)\coloneqq \sqrt{r_j} \rho_j \end{cases}\quad \forall j.
	\end{align*}
	There are again no boundary terms since we have a purely odd transformation.
	The Berezinian is given by $\prod_{j=1}^N r_j^{-1}$. Since $\psi_2$ is a linear transformation, this can also be computed directly.
	This cancels with the Jacobian from $\Psi_1$ up to a constant.
	Hence 
	\begin{align}\label{eq:Ifdef}
	I(f) = \frac{1}{\pi^N}\int_{(\R^+\times(0,2\pi))^N} \diff r \diff \theta \diff \bar\rho \diff \rho 
	\ f \circ \Psi_1\circ \Psi_2(r,\theta, \bar \rho,\rho),
	\end{align}
	where $\Psi_2 = \mathds{1} \times \psi_2$.
	After these transformations, we have $\bar z_j z_j+\bar \chi_j \chi_j = r_j^2+r_j \bar\rho_j\rho_j = (r_j + \tfrac12 \bar\rho_j\rho_j)^2$.
	We set 
	$\Psi_3(r,\theta,\bar\rho,\rho) =(r-\frac12 \bar \rho \rho,\theta,\bar\rho,\rho)	$.
	Hence $\Psi = \Psi_1 \circ \Psi_2 \circ \Psi_3 $ is the $\Psi$ from Eq. \eqref{eq:polarcoordinates}:
	\begin{align*}
	z_j &&\overset{\Psi_1}{\mapsto}&& r_j e^{i\theta_j}
	&&\overset{\Psi_2}{\mapsto}&& r_j e^{i\theta_j}
	&&\overset{\Psi_3}{\mapsto}&& \left(r_j-\tfrac 12 \bar \rho_j \rho_j\right)  e^{i\theta_j}, \\
	\chi_j &&\overset{\Psi_1}{\mapsto}&&\chi_j
	&&\overset{\Psi_2}{\mapsto}&& \sqrt{r_j} \rho_j
	&&\overset{\Psi_3}{\mapsto}&& \sqrt{r_j-\tfrac 12 \bar \rho_j \rho_j}\  \rho_j = \sqrt{r_j} \rho_j .
	\end{align*}
	We expand 
	$\tilde f = f\circ\Psi_1\circ\Psi_2\circ\Psi_3$ as follows
	\begin{align}
	\label{eq:expansionpsi3}
	%\tilde f (\mathbf{r},\theta,\bar\rho,\rho) =
	f\circ\Psi_1\circ\Psi_2(r,\theta,\bar\rho,\rho)
	= \tilde f (r+\tfrac{\bar\rho\rho}{2}, \theta,\bar\rho,\rho)
	= \mkern-10mu\sum_{\alpha\in\{0,1\}^N}   \mkern-10mu \left(\tfrac{\bar\rho\rho}{2}\right)^\alpha \partial_r^\alpha \tilde f(r,\theta, \bar\rho,\rho).
	\end{align}
	Note that we can set $\rho_j= 0$ and $\bar \rho_{j}=0$ for $\alpha_j = 1$ in   $\partial_r^\alpha  \tilde f$. We use the short-hand notation $\partial_r^\alpha \tilde f(r,\theta, \bar\rho,\rho)|_{ \bar\rho^\alpha = \rho^\alpha = 0}$. 
	Inserting \eqref{eq:expansionpsi3}  into the integral \eqref{eq:Ifdef}, we can integrate out the variables  $r^\alpha,$ $\rho^{\alpha}$ and $\bar\rho^{\alpha},$ as follows: 
	\begin{align}\label{eq:integrationbyparts}
	I(f) =& \tfrac{1}{\pi^N}\int_{(\R^+\times(0,2\pi))^N} \diff r \diff \theta \diff \bar\rho \diff \rho
	\sum_{\alpha\in\{0,1\}^N}2^{-|\alpha|}\left(\bar\rho\rho\right)^\alpha \partial_r^\alpha \tilde f(r,\theta, \bar\rho,\rho)\\ \nonumber
	=& \sum_{\alpha\in\{0,1\}^N}  \tfrac{1}{2^{|\alpha|}\pi^N}
	\int_{(\R^+)^{1-\alpha}\times(0,2\pi)^N}  \mkern-40mu (\diff r)^{1-\alpha} \diff \theta (\diff \bar\rho \diff \rho)^{1-\alpha}\tilde f(r,\theta, \bar\rho,\rho)|_{r^\alpha = \bar\rho^\alpha = \rho^\alpha = 0},
	\end{align}
	where in the second line we applied integration by parts in  $r^\alpha$
\[
\int_{(\R^{+})^{\alpha }} (dr)^{\alpha } \partial_{r}^{\alpha }\tilde{f} (r,\theta, \bar\rho,\rho)
	= (-1)^{\alpha } \tilde{f} (r,\theta, \bar\rho,\rho)\Big |_{|r^{\alpha }=0}
\]
and
\[
\int(\diff\bar\rho\diff\rho)^\alpha (\bar\rho\rho)^\alpha = (-1)^\alpha.
\]
	Note that $\tilde f (r,\theta, \bar\rho,\rho)|_{r^\alpha = \bar\rho^\alpha = \rho^\alpha = 0} = f\circ \Psi_\alpha$ is independent of $\theta_j$ for $\alpha_j =1 $ and we can integrate $\int (\diff \theta)^\alpha = (2\pi)^{|\alpha|}$. This proves the theorem.
\end{proof}

\subsection{Proof of Theorem \ref{theo:generalpolar}}
\begin{proof}[Proof of Theorem \ref{theo:generalpolar}]
	Applying Theorem \ref{theo:susyapproach} to $\mathcal{G}_\varepsilon(E,\tilde E)$ yields
	\begin{align*}
	\mathcal{G}_\varepsilon(E,\tilde E) = 
	\mathbb{E}\left[
	\int [\diff \Phi^* \diff \Phi ]\eto^{i \Phi^* (\mathbf{E} + i \varepsilon- \lambda V + \Delta)\Phi} \right].
	\end{align*}
	Note that we cannot interchange the average with the integral, since the average of the supersymmetric expression $\eto^{i\lambda \Phi^* V \Phi}$ may be ill-defined if  infinite moments are present. But after applying Theorem \ref{theo:polar} we get
	\begin{align*}
	\mathcal{G}_\varepsilon(E,\tilde E) = \hspace{-0,4cm}
	\sum_{\alpha\in\{0,1\}^\Lambda}  \hspace{-0,2cm}\pi^{-|1-\alpha|}
	\mathbb{E}\left[\int_{(\R^+ \times(0,2\pi))^{1-\alpha}}  \hspace{-1,4cm}
	(\diff r \diff \theta \diff \bar \rho \diff \rho)^{1-\alpha}
	\eto^{-i\lambda \sum_{j}V_j r_j^2}
	g\circ \Psi_\alpha (r,\theta,\bar \rho,\rho )
	\right],
	\end{align*}
	where $g(\Phi^*, \Phi) = \eto^{i \Phi^* (\mathbf{E} + i \varepsilon+ \Delta)\Phi} $.
	Now we can take the average inside the integral. 
	%\begin{align*}
	%\mathbb{E}[\eto^{-i \lambda \sum_{j} V_j r_j^2}] = 
	%\int_{\R^N} \mu(x) \eto^{i \lambda \sum_{j}x_j r_j^2} \diff x
	%= \hat {\mu}((\lambda r_j^2)_{j\in \Lambda}).
	%\end{align*}
	The same arguments hold for $	\mathbb{E}[|G_\Lambda(E+ i\varepsilon)_{jk}|^2]$.
\end{proof}

%Section 4: Application
\section{Applications to the Lloyd model} \label{sec:application}

\subsection{Proof of Proposition \ref{prop:Cauchyrepresentations}}

We will need the following well-known result for the proof of the proposition.

\begin{lemma}\label{lem:princvalue}
	Let $A\sim\text{Cauchy}(0,1)$ and $t\in\R$.
	Then $\mathbb{E}[\eto^{itA}] = \eto^{-|t|}$.
\end{lemma}

\begin{proof}
	Let $t\geq 0$. We take the principal value and apply the residue theorem.
	\begin{align*}
	\lim_{R\to\infty} \int_{[-R,R]}\frac{\eto^{it x}}{\pi(1+x^2)}\diff x
	=\lim_{R\to\infty}\left[ 2\pi i \frac{\eto^{it x}}{\pi(x+i)}\Big |_{x=i} - 
	\int_{\gamma}\frac{\eto^{it x}\diff x}{\pi(1+x^2)}\right]
	= \eto^{-t},
	\end{align*}
	where $\gamma(s) = R \eto^{is}$ for $s\in[0,\pi]$. The case $t <0$ follows analogously by 
	closing the contour from below.
\end{proof}

\begin{proof}[Proof of Proposition \ref{prop:Cauchyrepresentations}]
	Starting from the representation \eqref{eq:polarrepresentation} of Theorem \ref{theo:generalpolar}, we use Lemma \ref{lem:princvalue} to determine the Fourier transform 
	\begin{align*}
	\hat{ \mu}(\{\lambda r_j^2\}_{j\in\Lambda}) = \mathbb{E}[\eto^{i \lambda \sum_{j, k} T_{jk} W_k r_j^2}] 
	= \eto^{-\sum_{k} \lambda | \sum_{j} T_{jk} r_j^2|}.
	\end{align*}
	As long as $r_j\in \R$, this is well-defined and the integral remains finite for arbitrary correlation $T$. When $T_{jk} \geq 0$ for all $j,k$, we can drop the absolute value and obtain 
	\begin{align*}
	\hat{ \mu}(\{\lambda r_j^2\}_{j\in\Lambda}) = \eto^{-\sum_{k} \lambda  \sum_{j} T_{jk} r_j^2}
	= \tilde \mu \circ \Psi_\alpha,
	\end{align*}
	where $\tilde \mu(\Phi^*, \Phi) = \exp[-\sum_{k} \lambda \sum_{j} T_{jk} \Phi^*_j \Phi_j]$ is a smooth, integrable function in
	$\mathcal{A}_{2N,2N}(\C^N)$, which  can be transformed back to ordinary supersymmetric coordinates by Theorem \ref{theo:polar}.
	
	In the case of the toymodel, our function is continuous but only piecewise smooth. We partition the  integration domain into regions, where  our function is smooth. 
	In polar coordinates the regions \eqref{eq:Ibeta} become
	\begin{align*}
	\mathcal{I}_{++} &=\{0< \delta r_2 < r_1 <  \tfrac{r_2}{\delta}\} \times (0,\infty)^{\Lambda\backslash\{1,2\}} &=&
	\{r\in (0,\infty)^\Lambda : \delta r_2 < r_1 <  \tfrac{r_2}{\delta}\}, \\
	\mathcal{I}_{+-} &= \{0< \tfrac{r_2}{\delta}< r_1\} \times (0,\infty)^{\Lambda\backslash\{1,2\}}  &=& 
	\{r\in (0,\infty)^\Lambda :  r_1 >  \tfrac{r_2}{\delta}\}, \\
	\mathcal{I}_{-+} &= \{0< r_1 < \delta r_2\} \times (0,\infty)^{\Lambda\backslash\{1,2\}} &=&
	\{r\in (0,\infty)^\Lambda : r_1 < \delta r_2\}.
	\end{align*}
	Hence $(0,\infty)^\Lambda$ can be written as the disjoint union $ \mathcal{I}_{++} \cup \mathcal{I}_{+-} \cup \mathcal{I}_{-+} \cup \mathcal{N}$, where $\mathcal{N}$ is a set of measure $0$. 
	Using $T^\beta$ defined above, we can write 
	\begin{align*}
	I_1 = &\sum_{\alpha\in\{0,1\}^\Lambda}  \pi^{-|1-\alpha|} \int_{(\R^+\times(0,2\pi))^{1-\alpha}} \mkern-45mu
	(\diff r \diff \theta \diff \bar \rho \diff \rho)^{1-\alpha}
	\hat{ \mu}(\{\lambda r_j^2\}_{j\in\Lambda})|_{r^\alpha = 0} \ 
	g\circ \Psi_\alpha (r,\theta,\bar \rho,\rho)
	\\=& 
	\sum_{\beta}
	\sum_{\alpha\in\{0,1\}^\Lambda}  \pi^{-|1-\alpha|} \int_{(\R^+\times(0,2\pi))^{1-\alpha}} \mkern-40mu
	(\diff r \diff \theta \diff \bar \rho \diff \rho)^{1-\alpha} \chi(\mathcal{I}_\beta)|_{r^\alpha = 0} \\
	&\mkern150mu\times
	\eto^{-\lambda (\delta_{\alpha_1 0} r_1^2 T^\beta_1+ \delta_{\alpha_2 0} r_2^2 T^\beta_2)}
	h\circ \Psi_\alpha (r,\theta,\bar\rho,\rho),
	\end{align*}
	where $\beta \in \{++,+-,-+\}$ and $h (\Phi^*, \Phi) = g (\Phi^*,\Phi) \eto^{-\lambda \sum_{j\neq 1,2} \Phi_j^* \Phi_j}$ is independent of $\beta$. Finally, $\chi(\mathcal{I}_\beta)$ is the characteristic function of $\mathcal{I}_\beta$ and $r^\alpha = 0$ means $r_j = 0$ for $\alpha_j = 1$.
	
	To transform back we need to repeat the proof of Theorem \ref{theo:polar} on the different domains. Consider the integral
	\begin{align*}
	I_2 = \sum_{\beta }  \int_{{ \mathcal{I}}_\beta}[\diff \Phi^* \diff \Phi] \ h(\Phi^*,\Phi) \eto^{-\lambda \sum_{j=1}^2  T^\beta_j\Phi_j^* \Phi_j} ,
	\end{align*}
	where ${\mathcal{I}}_\beta $ are the corresponding subsets of $\C^\Lambda$ (cf. Eq. \eqref{eq:Ibeta}).
	We will show that inserting polar coordinates in $I_2$, we recover $I_1$ plus some correction terms.
	In each region, the integrated function is smooth and we can apply the first two transformations $\Psi_1$ and $\Psi_2$ from the proof of Theorem \ref{theo:polar} and obtain
	\begin{align*}
	I_2 = \frac{1}{\pi^{|\Lambda|}}\sum_{\beta}\int_{\mathcal{I}_\beta\times(0,2\pi)^{|\Lambda|}} \mkern-20mu \diff r \diff \theta \diff \bar \rho \diff \rho \   \eto^{-\lambda \sum_{j=1}^2  T^\beta_j (r_j+\frac 12 \bar \rho_j \rho_j)^2} h\circ \Psi_1 \circ \Psi_2(r,\theta,\bar\rho,\rho).
	\end{align*}
	Replacing as in Eq. \eqref{eq:expansionpsi3} the integrand by the Taylor-expansion of $\tilde f_\beta = 
	\eto^{-\lambda \sum_{j=1}^2  (T_\beta)_j r_j^2} \tilde h$, with $\tilde h = 
	h \circ \Psi _1 \circ \Psi_2 \circ \Psi_3$, we obtain
	\begin{align*}
	I_2 &=  \sum_{\alpha \in\{0,1\}^\Lambda} I_\alpha , \text{ where} \\
	I_\alpha &=  \frac{1}{\pi^{|\Lambda|}2^{|\alpha|}}\sum_{\beta} \int_{\mathcal{I}_\beta\times(0,2\pi)^{|\Lambda|}} \diff r \diff \theta \diff \bar \rho \diff \rho \   (\bar \rho \rho)^\alpha\  \partial_r^\alpha \tilde f_\beta(r,\theta,\bar \rho,\rho).
	\end{align*}	
	Applying now integration by parts as in Eq. \eqref{eq:integrationbyparts} generates additional boundary terms. More precisely we distinguish three cases.

\paragraph{Case 1: $\pmb{\alpha_1 = \alpha_2 = 0}$.} Here no derivatives in $r_1$ and $r_2$ appear and  $\mathcal{I}_\beta = \tilde {\mathcal{I}}_\beta\times (0,\infty)^{\Lambda \backslash\{1,2\}}$. Hence no additional terms arise and  
	\begin{align*}
	I_\alpha  &= \pi^{-|1-\alpha|}\sum_{\beta}
	\int_{(\R^+\times (0,2\pi))^{1-\alpha} } \mkern-45mu (\diff r \diff \theta \diff \bar \rho \diff \rho)^{1-\alpha} 
	\eto^{-\lambda \sum_{j=1}^2  (T_\beta)_j r_j^2}
	\chi(\tilde {\mathcal{I}}_\beta) \ h\circ \Psi_\alpha(r,\theta,\bar \rho,\rho).
	\end{align*}
\paragraph{Case 2: $\pmb{\alpha_1 =1, \alpha_2 = 0}$ (or vice versa).}
	Here additional boundary terms do appear but cancel since the function is continuous:
	\begin{align*}
	I_\alpha =& 
	\frac{1}{\pi^{|\Lambda|}2^{|\alpha|}}\sum_{\beta} \int_{\mathcal{I}_\beta\times(0,2\pi)^{|\Lambda|}} \mkern-20mu \diff r \diff \theta \diff \bar \rho \diff \rho  \ (\bar \rho \rho)^\alpha \ \partial_{r_1} \left[h^{(\alpha)}(r,\theta,\bar \rho,\rho) \eto^{-\lambda \sum_{j=1}^2 T^\beta_j r_j^2}\right] \\
	=&  \frac{1}{\pi^{|\Lambda|}2^{|\alpha|}} \int_{(\R^+)^{|\Lambda|-1}\times(0,2\pi)^{|\Lambda|}} \mkern-20mu \diff \hat r \diff \theta \diff \bar \rho \diff \rho \ (\bar \rho \rho)^\alpha 
	\left[
	h^{(\alpha)}\eto^{-\lambda \sum_{j=1}^2 T^{-+}_j r_j^2}\right]_{r_1= 0}^{r_1 = \delta r_2}
	\\ &
	+ \left[
	h^{(\alpha)}\eto^{-\lambda \sum_{j=1}^2 T^{++}_j r_j^2}\right]_{r_1= \delta r_2}^{r_1 =  r_2/ \delta} 
	+\left[
	h^{(\alpha)}\eto^{-\lambda \sum_{j=1}^2 T^{+-}_j r_j^2}\right]_{r_1= r_2/\delta}^{r_1 = \infty}
	\\
	%	&\left .  
	%	+ h^{(\alpha)}|_{r_1 = \delta r_2} \eto^{-\lambda (T^{-+}_2 +T^{-+}_1 \delta^2) r_2^2}
	%	- h^{(\alpha)}|_{r_1 = \delta r_2} \eto^{-\lambda (T^{++}_2 +T^{++}_1 \delta^2)r_2^2} \right.
	%	\\ & \left. \ 
	%	+ h^{(\alpha)}|_{r_1 =  r_2/\delta } \eto^{-\lambda (T^{++}_2 +T^{++}_1/ \delta^2) r_2^2}
	%	- h^{(\alpha)}|_{r_1 =  r_2/\delta} \eto^{-\lambda (T^{+-}_2 +T^{+-}_1/ \delta^2)r_2^2}
	%	\right] \\
	=&  -\frac{1}{\pi^{|\Lambda|}2^{|\alpha|}} \int_{(\R^+)^{|\Lambda|-1}\times(0,2\pi)^{|\Lambda|}} \mkern-20mu \diff \hat r \diff \theta \diff \bar \rho \diff \rho \ (\bar \rho \rho)^\alpha 
	h^{(\alpha)}|_{r_1 = 0}\eto^{-\lambda T^{-+}_2 r_2^2},
	%\pi^{-|1-\alpha|},
	%	\int_{(\R^+ \times (0,2\pi))^{1-\alpha} }  (\diff r \diff \theta \diff \bar \rho \diff \rho)^{1-\alpha} 
	%	\eto^{-\lambda  T^{-+}_2 r_2^2}
	%	h\circ \Psi_\alpha.
	\end{align*}
	where $\diff \hat r = \prod_{j\neq 1} \diff r_j$ and  $h^{(\alpha)} = \partial_{r}^{\hat \alpha}\tilde h $ and $\hat \alpha_j = \alpha_j$ for all $j\neq 1,2$, $\hat\alpha_1 = \hat{\alpha}_2= 0$.
	Note that in the second step all terms except the first one cancel because of continuity: $\sum_{j=1}^2 T^{-+}_j r_j^2|_{r_1 = \delta r_2} = \sum_{j=1}^2 T^{++}_j r_j^2|_{r_1 = \delta r_2}$ and $\sum_{j=1}^2 T^{++}_j r_j^2|_{r_1 = r_2/\delta} = \sum_{j=1}^2 T^{+-}_j r_j^2|_{r_1 = r_2/\delta}$. 
	We can apply now integration by parts for $r^{\hat\alpha}$ as before.	
	Note that for $r_1=0$ the sets $\mathcal{I}_{++} = \mathcal{I_{+-} = \emptyset}$ and we obtain only contributions from the set $\mathcal{I}_{-+}= \{r_2 \in \R^+\}$ which is the same as writing $\sum_{\beta } \chi(\mathcal{I}_\beta)|_{r_1 = 0}$.

	\paragraph{Case 3: $\pmb{\alpha_1 = \alpha_2 = 1}$.} Here we obtain additional boundary terms which do not cancel. Applying integration by parts in $r_1$, we need to evaluate 
	\begin{align*}
	\partial_{r_2} [h^{(\alpha)} \eto^{-\lambda \sum_{j=1}^2 (T_\beta)_j r_j^2}] = (\partial_{r_2} h^{(\alpha)}-2 \lambda T^\beta_2 r_2 h^{(\alpha)})\eto^{-\lambda \sum_{j=1}^2 (T_\beta)_j r_j^2}
	\end{align*}
	on the different boundaries. The contributions of $\partial_{r_2} h^{(\alpha)} \eto^{-\lambda\sum_{j=1}^2 (T_\beta)_j r_j^2}$ cancel as above by continuity except for the term at $r_1=0$. The contributions from the second summand remain. Precisely we get
	\begin{align*}
	I_\alpha =& 
	\frac{1}{\pi^{|\Lambda|}2^{|\alpha|}}\sum_{\beta} \int_{\mathcal{I}_\beta\times(0,2\pi)^{|\Lambda|}} \diff r \diff \theta \diff \bar \rho \diff \rho \ (\bar \rho \rho)^\alpha \partial_{r_1}\partial_{r_2} \left[h^{(\alpha)} \eto^{-\lambda \sum_{j=1}^2 (T_\beta)_j r_j^2}\right] \\
	=&  \frac{1}{\pi^{|\Lambda|}2^{|\alpha|}} \int_{(\R^+)^{|\Lambda|-1}\times(0,2\pi)^{|\Lambda|}} \mkern-40mu\diff \hat r \diff \theta \diff \bar \rho \diff \rho \ (\bar \rho \rho)^\alpha  \partial_{r_2}\left[
	- h^{(\alpha)}\eto^{-\lambda T^{-+}_2 r_2^2}
	\right ]_{r_1= 0} + R_\alpha(h),
	\end{align*}
	where $R_\alpha (h)$ is  defined below in Eq. \eqref{eq:Ralpha(h)}. In the first integral, we can apply integration by parts in $r_2$ and $r^{\hat{\alpha}}$ as before and the result is independent of $\beta$. It remains to consider 
	\begin{align}\label{eq:Ralpha(h)}
	&R_\alpha(h) =   \frac{1}{\pi^{|\Lambda|}2^{|\alpha|}} \int_{(\R^+)^{|\Lambda|-1}\times(0,2\pi)^{|\Lambda|}} \mkern-20mu\diff \hat r \diff \theta \diff \bar \rho \diff \rho \ (\bar \rho \rho)^\alpha \  2 \lambda r_2 
	\\
	&\times\!\left [ h^{(\alpha)}|_{r_1 = \delta r_2}  (T^{++}_2\!\! -T^{-+}_2) \eto^{-\lambda (1-\delta^4) r_2^2} + h^{(\alpha)}|_{r_1 =  \frac{r_2}{\delta} }  (T^{+-}_2\!\! -T^{++}_2) \eto^{-\lambda (1-\delta^4) \delta^{-2} r_2^2}
	\right]. \nonumber
	\end{align}
	Here, we can integrate over $r^{\hat \alpha},$ $\theta^{\hat \alpha },$ $\rho^{\alpha}$ and $\bar \rho^{\alpha }$ first as follows:
\begin{align*}
	&\int_{(\R^+\times (0,2\pi))^{\hat\alpha}} \mkern-50mu (\diff r \diff \theta )^{\hat \alpha }(\diff \bar \rho \diff \rho)^{\alpha}
	 (\bar \rho \rho)^\alpha  h^{(\alpha)} \\
&\qquad \qquad = (-2\pi)^{\hat \alpha} (-1)^{\alpha} h|_{r^{\hat \alpha} = \bar \rho^\alpha=  \rho^\alpha= 0}
= (2\pi)^{\hat \alpha} h|_{r^{\hat \alpha} = \bar \rho^\alpha=  \rho^\alpha= 0}.
\end{align*}
where we used $|\alpha|=|\hat\alpha |+2.$ The resulting integral is:
	\begin{align*}
	&R_\alpha(h) = - \pi^{-|1-\hat \alpha|}
	\int_{(\R^+\times (0,2\pi))^{1-\hat\alpha}\times \R^+ \times (0,2\pi)^2 } \mkern-50mu (\diff r \diff \theta )^{1- \hat \alpha }(\diff \bar \rho \diff \rho)^{1- \alpha} \diff r_2 \diff \theta_1 \diff \theta_2 \ \lambda r_2 \\
	&\times \left[
	\tilde h|_{r^{\hat \alpha} = \bar \rho^\alpha=  \rho^\alpha= 0, r_1 = \delta r_2 }  \delta^2 \eto^{-\lambda (1-\delta^4)r_2^2} +	 \ \tilde h|_{r^{\hat \alpha} = \bar \rho^\alpha = \rho ^ \alpha = 0, r_1 = r_2/\delta  }  \eto^{-\lambda (1-\delta^4)\delta^{-2} r_2^2} \right].
	\end{align*}
	After rescaling the second term $r_2 \mapsto \delta^2 r_2$, we obtain
	\begin{align*}
	R_{\alpha } (h) =& -
	\pi^{-|1-\hat \alpha|}
	\int_{(\R^+\times (0,2\pi))^{1-\hat\alpha}\times \R^+ \times (0,2\pi)^2 } \mkern-50mu (\diff r \diff \theta )^{1- \hat \alpha }(\diff \bar \rho \diff \rho)^{1- \alpha} \diff r_2 \diff \theta_1 \diff \theta_2 \\
	&\times 
	\lambda r_2 \delta^2  \ \tilde h|_{r^{\hat \alpha}= \bar \rho^\alpha = \rho^\alpha = 0, r_1 = \delta r_2 }  \left[\eto^{-\lambda (1-\delta^4)r_2^2} +\delta^2 \eto^{-\lambda (1-\delta^4)\delta^2 r_2^2} \right].
	\end{align*}
As a result we proved that
$I_{2}-I_{1}= \sum_{\hat\alpha} R_{\alpha } (h).$ Note that in above representation for $R_{\alpha } (h)$ we can transform the
variables of $\Lambda \backslash \{1,2\}$ back to flat coordinates by Theorem \ref{theo:polar}.  This yelds \eqref{eq:R(h)}
and thus completes the proof.
\end{proof}

\subsection{Proof of Theorem \ref{theo:mainresult}}
\begin{proof}[Proof of Theorem \ref{theo:mainresult}]
	We start from the result of Propostion \ref{prop:Cauchyrepresentations}.
	
	In both models, the classical and the positive correlated one, we have $T_{jk}\geq 0$ and $\sum_k T_{jk} >0$, hence the body of $\lambda \sum_{j} T_{jk} \Phi^*_j \Phi_j$ is strictly positive except on a set of measure $0$. We end up with
	\begin{align*}
	\mathcal{G}_\varepsilon(E,\tilde E) =
	% \mathbb{E}\left[ \frac{\det (E+ i\varepsilon - H)}{\det (\tilde E + i \varepsilon- H)}\right] = 
	\int [\diff \Phi^* \diff \Phi ]\eto^{i \Phi^* (\hat E + i \varepsilon + i \lambda \hat T+ \Delta)\Phi} ,
	\end{align*}
	where we can take the average $\varepsilon \to 0$ and go back to the original representation. 
\end{proof}

\subsection{Proof of Theorem \ref{theo:toymodel}}

\begin{proof}[Proof of Theorem \ref{theo:toymodel}]
	Using Eq. \eqref{eq:gree-repr} and the result of Proposition \ref{prop:Cauchyrepresentations}, we obtain
	\begin{align*}
	\mathbb{E}[\trace G_{\Lambda }(E+i\varepsilon )]&= \mathbb{E}\Big[ \int [\diff \Phi^* \diff \Phi] \eto^{i\Phi^* (E+i\varepsilon-\lambda V + \Delta) \Phi } \sum_{j\in\Lambda} |z_j|^2\Big] \\
	&= I_{++}+I_{+-}+I_{-+}+ R(h)
	\end{align*}
	where for $\beta = (++), (+-)$ or $(-+)$ we have
	\begin{align*}
	I_\beta &= \int_{\mathcal{I}_\beta} [\diff \Phi^* \diff \Phi] \eto^{i\Phi^* (E+i\varepsilon + \Delta) \Phi } \sum_{j\in\Lambda} |z_j|^2 \eto^{-\lambda(T_{1}^{\beta }\Phi_1^{*}\Phi_1 + T_{2}^{\beta } \Phi_2^{*}\Phi_2+ \sum_{k\neq 1,2} \Phi_k^* \Phi_k)},
	\end{align*}
	and for $h(\Phi^*,\Phi)=\sum_{j} |z_j|^2 \eto^{-\lambda \sum_{j\neq 1,2}\Phi^*_j \Phi_j}\eto^{i \Phi^* ( E - i \varepsilon + \Delta)\Phi}$ the remainder $R(h)$ is defined in Eq. \eqref{eq:R(h)}. 
	
	We will show that the main contribution comes from  $\mathcal{I}_{++}$ and indeed
	\[
	\text{ body } ( T_{1}^{++} \Phi_1^{*}\Phi_1 + T_{2}^{++}\Phi_2^{*}\Phi_2)= (1-\delta^{2}) [|z_{1}|^{2}+|z_{2}|^{2}]>0\quad \forall (z_{1},z_{2})\neq (0,0).
	\]
	In the following we show that $I_{+-}$ and $I_{-+},$ as well as $R(h)$ are small in terms of $\delta$.
	\paragraph{Analysis of the $\pmb{I_{\beta }}$ terms.}
	Integrating out the Grassmann variables, we obtain for all $\beta$
	\begin{align*}
	I_\beta = \int_{\mathcal{I}_\beta} \diff \bar z \diff z  \sum_{j\in\Lambda} |z_j|^2 \det\left[ \tfrac{C_\beta+\varepsilon }{2\pi} \right] \eto^{-\bar z ( C_\beta+\varepsilon ) z } ,
	\end{align*}
	where   $C_{\beta }$ has the block structure
	\begin{align}
	C_\beta& = 
	\begin{pmatrix}
	A_\beta & -iD \\
	-iD^T & B
	\end{pmatrix},
	\quad  A_\beta \coloneqq A_0+ \lambda \diag T^\beta, \quad A_0 \coloneqq -i(E+\Delta)|_{\{1,2\}}\nonumber\\
	\label{eq:defblockmatrix}
	%\begin{pmatrix}
	%\lambda T_{1}^{\beta } -i (E-2d) & -i \\
	%-i & \lambda T_{2}^{\beta} - i (E-2d)
	%\end{pmatrix}\\
	B& \coloneqq (\lambda - i (E+\Delta))_{|\Lambda \setminus \{1,2 \}},\quad D^T\coloneqq (d_{1},d_{2}),
	\end{align}
	and we defined the vectors $d_{1},d_{2}\in \R^{\Lambda \setminus \{1,2 \}}$ as
	$d_{1} (j)=\delta_{|i_{1}-j|,1},$ $d_{2} (j)=\delta_{|i_{2}-j|,1},$ where $i_{1},i_{2}$ are the positions of $1,2.$
	Note that the blocks $B$ and $D$ are independent of $\beta$ and $\re B>0.$  
	On the contrary $\re C_{\beta}>0$ holds only for $\beta = (+,+).$ 
	We set then $\varepsilon =0$ in our formulas and  reorganize   $I_{++}+I_{+-}+I_{-+}$  as follows
	\begin{align*}
	&[I_{++}+I_{+-}+I_{-+}]_{|\varepsilon =0} = \int\diff \bar z \diff z  \sum_{j\in\Lambda} |z_j|^2 \det\left[ \tfrac{C_{++}}{2\pi} \right] \eto^{-\bar z C_{++} z } \\
	&\qquad +
	\int_{\mathcal{I}_{+-}}  \diff \bar z \diff z  \sum_{j\in\Lambda} |z_j|^2\left( \det\left[ \tfrac{C_{+-}}{2\pi} \right] \eto^{-\bar z C_{+-} z }- \det\left[ \tfrac{C_{++}}{2\pi} \right] \eto^{-\bar z C_{++} z }\right)\\
	&\qquad +
	\int_{\mathcal{I}_{-+}}  \diff \bar z \diff z  \sum_{j\in\Lambda} |z_j|^2\left( \det\left[ \tfrac{C_{-+}}{2\pi} \right] \eto^{-\bar z C_{-+} z }- \det\left[ \tfrac{C_{++}}{2\pi} \right] \eto^{-\bar z C_{++} z }\right)\\
	&\qquad = \trace C_{++}^{-1} + \int_{\mathcal{I}_{+-}} (\cdots) + \int_{\mathcal{I}_{-+}}(\cdots)=
	\trace C_{++}^{-1} \left (  1+ \mathcal{E}_{+-}+  \mathcal{E}_{-+} \right )
	\end{align*}
	To estimate $\mathcal{E}_{+-}$ and $\mathcal{E}_{-+}$, we integrate over the variables $w = (z_j)_{j\in\Lambda, j\neq 1,2}$ exactly. 
	We define $z= (\hat z, w), \hat z = (z_1, z_2).$ Then 
	\begin{align*}
	\bar z C_{\beta } z&=  \bar w B w -i \bar w D^{t} \hat{z}-i \overline{ \hat z} D w+ \overline{ \hat z} A_{\beta } \hat z ,\\
	\sum_{j\in\Lambda} |z_j|^2&=\overline{ \hat z}\hat z+ \sum_{l\in \Lambda \setminus \{1,2 \}}  |w_l|^2.
	\end{align*}
	Integrating over $w$ we  get
	\begin{align*}
	&\int \diff \bar w \diff w \det\left[ \tfrac{B}{2\pi} \right] \eto^{-\bar w B w } \eto^{-i\bar w D^{t} \hat{z}-i \overline{ \hat z} D w}
	(\bar w w+ \overline{ \hat z}\hat z)\\
	&=  \eto^{- \overline{ \hat z} DB^{-1}D^{t} \hat{z} }
	(\trace B^{-1} - \overline{ \hat z} D B^{-2}D^{t} \hat{z}+ \overline{ \hat z}\hat z)=
	\eto^{- \overline{ \hat z} DB^{-1}D^{t} \hat{z} }
	(\trace B^{-1} + \overline{ \hat z} M \hat{z} ),
	\end{align*}
	where  we defined  $M \coloneqq 1 - DB^{-2}D^T.$
	Then for $\beta= (+-),(-+)$ and  $\beta'=\beta$ or $\beta'= (++)$ we have
	\begin{align*}
	&\int_{\mathcal{I}_\beta} \diff \bar z \diff  z \det\left[ \tfrac{C_{\beta'}}{2\pi} \right] \eto^{-\bar z C_{\beta'} z }  \sum_{j\in \Lambda} |z_j|^2 \\& \qquad = 
	\int_{\mathcal{I}_\beta} \diff \overline{ \hat z} \diff \hat z \det  \left[ \tfrac{S_{\beta'}}{2\pi} \right] \eto^{-\overline {\hat z} S_{\beta'} \hat z } 
	\left(
	\trace B^{-1} + \overline{\hat z } M \hat z
	\right),
	\end{align*}
	where $S_{\beta'} = A_{\beta'}+D B^{-1} D^T$ is the Schur complement of the $2\times 2$ block of $C_{\beta'}$
	corresponding to $1,2$. We
	also used $\det C_{\beta'}= \det B \det S_{\beta'}$.  
	%Note that $S_{++}$ is invertible, also for $\varepsilon =0$ since $\hat T_{++}|_{12} >0$. %Note that $\det C_j^{-1}= \det B \det S_j^{-1}$. % and $\trace C_j = \trace B^{-1} + \trace S_j M$.
	We consider now the error term $\mathcal{E}_{-+}.$ The error term $\mathcal{E}_{+-}$ works analogously.
	From the results above we get
	\begin{align*}
	&\mathcal{E}_{-+} =\frac{1}{\trace C_{++}^{-1}}\int_{\mathcal{I}_{-+}}\diff \bar z \diff z  \sum_{j\in\Lambda} |z_j|^2\left( \det\left[ \tfrac{C_{-+}}{2\pi} \right] \eto^{-\bar z C_{-+} z }- \det\left[ \tfrac{C_{++}}{2\pi} \right] \eto^{-\bar z C_{++} z }\right) \\
	&= \int_{|z_1| <\delta |z_2|} \diff\overline{\hat z} \diff \hat z \det \left[\tfrac{S_{++}}{2\pi}\right]
	\eto^{-\overline {\hat z } S_{++} \hat z}
	\, \tfrac{\trace B^{-1} + \overline{\hat z } M \hat z}{\trace C_{++}^{-1}} 
	%\\ &\qquad\times
	\left(\eto^{- \overline{\hat z } X \hat z}\det (1 + S_{++}^{-1}X) -1 \right),
	\end{align*}
	where we used  $S_{++}^{-1}$  invertible and we defined
	\begin{align*}
	X\coloneqq A_{-+}- A_{++}=2\lambda 
	\begin{pmatrix}
	-1  & 0 \\ 0 &  \delta^2
	\end{pmatrix}, \, \text{ hence } \, 
	\overline{\hat z }X \hat z = 2\lambda  (\delta^2 |z_2|^2 - |z_1|^2) >0 .
	\end{align*}
	Now we change the  coordinate $z_1$ to $v = z_1 z_2^{-1} \delta^{-1}.$
	As a short-hand notation write $S = S_{++}$. We have
	\[
	\overline {\hat z } S \hat z= |z_{2}|^{2} (\mathbf{v}^* S\mathbf{v}),\qquad  \overline{\hat z } M \hat z=
	|z_2|^2 \mathbf{v}^* M \mathbf{v},
	\qquad  \overline {\hat z } X \hat z= |z_{2}|^{2} (\mathbf{v}^* X\mathbf{v}),
	\]
	where  $\mathbf{v}= (\delta v,1)^t$ and $\mathbf{v}^* = (\delta\bar v,1 )$.
	Note that $\re S>0$ and  
	\begin{align}\label{eq:Xestimate}
	(\mathbf{v}^* X\mathbf{v})= 2\lambda \delta^{2} (1-|v|^{2})\geq 0,
	\end{align}
	therefore we can  perform the integral over $z_{2}$ exactly
	\begin{align*}
	\mathcal{E}_{-+} =&  \det \left[\tfrac{S}{2\pi}\right]
	\int_{|v|<1} 
	\diff \bar z_2 \diff z_2 \diff \bar v \diff v \  \delta^2 |z_2|^2 \eto^{-|z_2|^2 \mathbf{v}^* S\mathbf{v}}
	\tfrac{\trace B^{-1} + |z_2|^2 \mathbf{v}^* M \mathbf{v}}{\trace C_{++}^{-1}}\\
	&\times \left(\eto^{-|z_2|^2 \mathbf{v}^*X \mathbf{v}}\det (1 +  S^{-1}X)-1 \right)\\
	=& 
	\delta^2\int_{|v|<1} 
	\frac{\diff \bar v \diff v}{2\pi } \left[
	\frac{\trace B^{-1}}{\trace C_{++}^{-1}}  \left(\frac{\det (S + X)}{(\mathbf{v}^* (S+ X)\mathbf{v})^2}-
	\frac{\det S}{(\mathbf{v}^* S\mathbf{v})^2}\right)\right. \\
	& \, + \left.
	\frac{2 \mathbf{v}^* M \mathbf{v}}{\trace C_{++}^{-1}}  \left(\frac{\det (S + X)}{(\mathbf{v}^* (S+X)\mathbf{v})^3}-
	\frac{\det S}{(\mathbf{v}^* S\mathbf{v})^3}\right)
	\right]\\
	=& \delta^2\int_{|v|<1} 
	\frac{\diff \bar v \diff v}{2\pi }
	\left(\frac{\det (S + X)}{(\mathbf{v}^* (S+ X)\mathbf{v})^2}-
	\frac{\det S}{(\mathbf{v}^* S\mathbf{v})^2}\right)  + O (|\Lambda |^{-1}),
	\end{align*}
	where we applied Lemma \ref{lem:matrixentries} below and 
	\begin{align}\label{eq:traceBC}
	\frac{\trace B^{-1}}{\trace C_{++}^{-1}}=1- \frac{\trace S^{-1}_{++}M}{\trace C^{-1}_{++}}= 1+O (|\Lambda |^{-1}).
	\end{align}
	Applying Lemma \ref{lem:matrixentries} again, together with Eq. \eqref{eq:Xestimate} we get
	\begin{align*}
	&\frac{\det (S + X)}{(\mathbf{v}^* (S+ X)\mathbf{v})^2}-
	\frac{\det S}{(\mathbf{v}^* S\mathbf{v})^2}  =
	-\frac{(\mathbf{v}^*X\mathbf{v})\det S}{(\mathbf{v}^* (S+ X)\mathbf{v})^2(\mathbf{v}^* S\mathbf{v})}
	\left[2+\frac{(\mathbf{v}^*X\mathbf{v})}{(\mathbf{v}^* S\mathbf{v})} \right]\\
	&+ \frac{X_{11}S_{22}+X_{22}S_{11}+X_{11}X_{22}}{(\mathbf{v}^* (S+ X)\mathbf{v})^2}= \mathcal{O}\Big ( (1+\frac{1}{\lambda^{2}})
	\left[1+\delta^{2}(1+\frac{1}{\lambda^{2}}) \right]\Big ).
	\end{align*}
	\paragraph{Analysis of $\pmb{R(h)}$.}
	Note that we can set $\varepsilon = 0$ in $R(h)$. By the notations in Eq. \eqref{eq:defblockmatrix}, we can write
	\begin{align*}
	R(h) =&- \tfrac{1}{\pi^2} \int_{\R^+ \times (0,2\pi)^2} \mkern-20mu \diff r_2 \diff \theta_1 \diff \theta_2 [\diff \hat{\Phi}^* \diff \hat \Phi]  \lambda r_2 \delta^2 \left[\eto^{-\lambda (1-\delta^4) r_2^2} +\delta^2 \eto^{-\lambda (1-\delta^4)\delta^2 r_2^2} \right]
	\\&\times
	\left((1+\delta^2) r_2^2 +\sum_{j} |w_j|^2\right) \eto^{-\hat{\Phi}^* B \hat \Phi }
	\eto^{ir_2 (\bar w D^T v_\theta+ \bar v_\theta D w)}
	\eto^{-r_2^2 \bar v_\theta A_0 v_\theta},
	\end{align*}
	where $v_\theta^t = (\eto^{i \theta_1}\delta, \eto^{i \theta_2})$.
	Integrating over the Grassmann variables, we obtain
	\begin{align*}
	R(h) =& - \tfrac{1}{\pi^2}  \int_{\R^+ \times (0,2\pi)^2} \mkern-20mu \diff r_2 \diff \theta_1 \diff \theta_2\diff \bar w \diff w \lambda r_2 \delta^2 \left[\eto^{-\lambda (1-\delta^4)r_2^2} +\delta^2 \eto^{-\lambda (1-\delta^4)\delta^2 r_2^2} \right]
	\\&\times
	\left((1+\delta^2) r_2^2 +\sum_{j} |w_j|^2\right) \det\left[\tfrac{B}{2\pi}\right]\eto^{-\bar w B w }
	\eto^{ir_2 (\bar w D^T v_\theta+ \bar v_\theta D w)}
	\eto^{- r_2^2\bar v_\theta A_0 v_\theta},
	\end{align*}
	Define $S_0 = A_0 + DB^{-1}D^T$. Integrating over $w$ and $r_2$, we obtain
	\begin{align*}
	\frac{R(h)}{\trace C_{++}^{-1}} = & - \tfrac{1}{\pi^2}\tfrac{1}{\trace C_{++}^{-1}}\int_{\R^+ \times (0,2 \pi)^2  } \mkern-30mu \diff r_2 \diff \theta_1 \diff \theta_2 \lambda r_2 \delta^2 \left[\eto^{-\lambda (1-\delta^4)r_2^2} +\delta^2 \eto^{-\lambda (1-\delta^4)\delta^2 r_2^2} \right]
	\\&\times
	\left( \bar v_\theta M v_\theta r_2^2 +\trace B^{-1}\right) \eto^{-r_2^2 \bar v_\theta S_0 v_\theta} \\
	=&- \tfrac{1}{\pi^2} \tfrac {\trace B^{-1}}{\trace{C_{++}^{-1}}}\int_{ (0,2 \pi)^2 } \mkern-20mu \diff \theta_1 \diff \theta_2 \tfrac{\lambda  \delta^2 }{2}\left[
	\tfrac{1}{\lambda (1-\delta^4)+ \bar v_\theta S_0 v_\theta }  + \tfrac{\delta^2}{\lambda \delta^2 (1-\delta^4)+ \bar v_\theta S_0 v_\theta }
	\right]
	\\& - \tfrac{1}{\pi^2}\int_{(0,2\pi)^2} \mkern-20mu \diff \theta_1 \diff \theta_2 \tfrac{\lambda \delta^2}{2} \tfrac{\bar v_\theta M v_\theta }{\trace{C_{++}^{-1}}}
	\left[
	\tfrac{1}{(\lambda (1-\delta^4)+ \bar v_\theta S_0 v_\theta )^2}  + \tfrac{\delta^2}{(\lambda \delta^2 (1-\delta^4)+ \bar v_\theta S_0 v_\theta)^2 }
	\right].
	\end{align*}
	Similar to the estimates above, we insert absolute values and use Lemma  \ref{lem:matrixentries} and Eq. \eqref{eq:traceBC} to bound the first term by $\delta^2 (1+ \mathcal{O}(|\Lambda|^{-1}))$ and the second one by $\delta^2\mathcal{O}(\lambda^{-1}|\Lambda|^{-1})$.
\end{proof}

\begin{lemma}\label{lem:matrixentries}
	Let  %$\lambda_0 = \max (6 |E-2d|, \sqrt{12}(2d-1))$,
	$\eta >0$ and $\mu_{\lambda } = \frac{\lambda \eta}{\lambda + 4 d \eto^\eta}$. Let $B, M,  C_{++}$ and $S_{++}$ be
	the matrices in the proof above. Set $0<\delta\leq \frac{1}{2}. $ Then
	\begin{enumerate}[(i)]
		\item \label{lem:1} $|B^{-1}_{ij}| \leq \frac{2}{\lambda} \eto^{-\mu_{\lambda }|i-j|}$ and
		$\re (f^{*}B^{-1}f)\geq \frac{\lambda }{\lambda^{2}+ (4d)^{2}}\|f\|^2$ $\forall f\in \C^{\Lambda \setminus\{1,2 \}}$
		%\item \label{lem:2}$|M_{ij}| \leq \delta_{ij} + \frac{4 (2d-1)^2}{\lambda^2 \mu^d_{\lambda }}$.				
		\item \label{lem:6}$|\trace C_{++}^{-1} | \geq \frac{ |\Lambda| \lambda }{K(\lambda+1)^{2}}.$
		\item \label{lem:3} $\re (f^{*}S_{++}f)\geq \frac{\lambda }{2 }\|f\|^2$ $\forall f\in \C^{\Lambda \setminus\{1,2 \}}.$
		Moreover
		
		$ |(S_{++})_{jk}| \leq  K (\lambda+\frac{1}{\lambda })$ for all $j,k= 1,2$ .
		%	\item \label{lem:4}$|(S_{++})_{12}| = | (S_{++})_{21})| \leq 2d $ for all $\lambda \geq \lambda_0$.
		%	\item \label{lem:5}$|\det S_{++}| \leq 3 \lambda^2 $ for all $\lambda \geq \lambda_0$.
	\end{enumerate}
\end{lemma}

\begin{proof}%\hspace{2cm}
	$(i)$ We have $B = i (-\Delta_{|\Lambda \setminus \{1,2 \}}- (E+i\lambda))$. The upper bound follows by  Combes-Thomas \cite{AizWar}[Sect 10.3]. For the lower bound note that
	\[
	f^{*}\re B^{-1} f = \lambda \|B^{-1} f\|^2
	\]
	Moreover $\| B g \|^2 =   \lambda^2 \|g\|^2+ g^{*}(E+\Delta_{|\Lambda \setminus \{1,2 \}} )g \leq (\lambda^{2}+ (4d)^{2})  \|g\|^2.$
	The result follows setting 
	$g= B^{-1} f$.
	\\
	$(ii)$ As in $(i)$ above $	f^{*}\re C_{++}^{-1} f \geq \lambda (1-\delta^{2})\|C_{++}^{-1}f\|^2.$
	We can write $C_{++}=\lambda -\lambda \delta^{2}\mathds{1}_{1,2}-i (E+\Delta),$ where $\mathds{1}_{1,2}$ is the diagonal matrix
	$(\mathds{1}_{1,2})_{ij}=\delta_{ij}[\delta_{ji_{1}}+\delta_{ji_{2}}]. $ Hence 
	\[
	C_{++}^{*}C_{++}= (\lambda -\lambda \delta^{2}\mathds{1}_{1,2})^{2}+  (E+ \Delta)^2+i\lambda \delta^{2} [\mathds{1}_{1,2},\Delta ].
	\]
	The result follows by inserting this decomposition in $\|C_{++}g\|^{2}$ for $g=C_{++}^{-1}f.$ 
	\\
	$(iii)$ Using $(i)$ we have 
	\[
	\re f^{*}Sf=\lambda (1-\delta^{2})\|f\|^{2} +\re f^{*}DB^{-1}D^{t}f\geq \lambda (1-\delta^{2})\|f\|^{2}.
	\]
	The upper bound follows from  $(i)$ too.
\end{proof}

\begin{appendix}
	
\section{Super analysis}\label{app:susy}

We collect here only a minimal set of definitions for our purpose. 
For details, see \cite{Berezin1987, Varadarajan2004,Wegner2016,DeWitt1992}.

\subsection{Basic definitions}

Let $q\in\N$.
Let $\mathcal{A}=\mathcal{A}_q = \mathcal{A}[\chi_1,\dots,\chi_q]$
be the Grassmann algebra over $\C$ generated by $\chi_1, \dots,\chi_q$, i.e. 
\begin{align*}
\mathcal{A} = \oplus_{i=0}^q V^i,
\end{align*}
where $V$ is the complex vector space with basis $(\chi_1,\dots,\chi_q)$, 
$V^0 = \C$ and $V^{j} = V^{j-1} \wedge V$
for $j\geq 2$ with  the anticommutative product  $\wedge$  
\begin{align*}
\chi_i \wedge \chi_j = - \chi_j \wedge \chi_i.
\end{align*}
As a short hand notation, we write in the following 
$\chi_i  \chi_j = \chi_i \wedge\chi_j$ and for
$I\subset \{1,\dots,q\}$ denote $\chi^I = \prod_{j\in I} \chi_j$ 
the \emph{ordered} product of the $\chi_j$ with $j\in I$.
Then each $a\in\mathcal{A}$ has the form
\begin{align}\label{eq:grassmannelement}
a = \sum_{I\in\mathcal{P}(q)} a_I \chi^I,
\end{align}
where $\mathcal{P}(q)$ is the power set of $\{1,\dots, q\}$ and $a_I\in\C$ 
for all $I\in\mathcal{P}(q)$.
We distinguish even and odd elements  
$\mathcal{A}=\mathcal{A}^0 \oplus\mathcal{A}^1$, where
\begin{align*}
\mathcal{A}^0 = \oplus_{i=0}^{\lfloor q/2\rfloor} V^{2i}, 
\quad 
\mathcal{A}^1 = \oplus_{i=0}^{\lfloor q/2\rfloor} V^{2i+1}.
\end{align*}
The parity operator $p$ for homogeneous 
(i.e. purely even, resp. purely odd) elements is defined by
\begin{align*}
p(a) = 
\begin{cases}
0 & \text{if } a\in\mathcal{A}^0,\\
1 & \text{if } a\in\mathcal{A}^1.
\end{cases}
\end{align*}
Note that even elements commute with all elements in $\mathcal{A}$ 
and two odd elements anticommute.
For an even element $a\in\mathcal{A}^0$, we write $a = b_a+ n_a$,
where $n_a$ is the nilpotent part
and $b_a = a_\emptyset \in\C$ is called the body of $a$.

Let $U\subset \R$ open. For any function $f\in C^{\infty}(U)$, we define 
\begin{align}
\label{eq:upgradefunc}
\begin{split}
f:\mathcal{A}^0 &\to \mathcal{A}^0\\
a & \mapsto f(a) = f(b_a+ n_a) 
= \sum_{k=0}^\infty  \frac{1}{k!}f^{(k)} (b_a) n_a^k
\end{split}
\end{align}
via its Taylor expansion. Note that the sum above is always finite.

\subsection{Differentiation}

Let $I' \subset I$. 
We define the signs $\sigma_l (I,I')$ and $\sigma_r (I,I')$ via
\begin{align*}
\chi^I = \sigma_l (I,I') \chi^{I'} \chi^{I\backslash I'} 
\quad 
\chi^I = \sigma_r (I,I') \chi^{I \backslash I'} \chi^{I'} .
\end{align*}
Then the left- resp. right-derivative of an element $a$ 
of the form \eqref{eq:grassmannelement} is defined as
\begin{align*}
\overrightarrow{\frac{\partial}{\partial \chi_j}} a 
\coloneqq \sum_{I\in\mathcal{P}(q): j \in I} 
a_I\  \sigma_l (I,\{j\}) \ \chi^{I\backslash\{j\}}, \\
a \overleftarrow{\frac{\partial}{\partial \chi_j}} 
\coloneqq \sum_{I\in\mathcal{P}(q): j \in I} 
a_I\  \sigma_r (I,\{j\}) \ \chi^{I\backslash\{j\}}.
\end{align*}

\subsection{Integration}

The integration over a subset of (odd) generators $\chi_j, j\in I$ is defined by
\begin{align*}
\int \diff \chi^I a 
\coloneqq \left(\overrightarrow{\frac{\partial}{\partial \chi}}\right)^I a 
= \sum_{J\in\mathcal{P}(q): I \subset J} 
a_J\  \sigma_l (J,I) \ \chi^{J\backslash I},
\end{align*}
where $a$ has the form \eqref{eq:grassmannelement} 
and $\diff \chi^I = \prod_{j\in I} \diff \chi_j$ is again a ordered product.
Note that the one forms $\diff \chi_i$ are anticommutative objects and e.g.
$\int \diff \chi_i \diff \chi_j \chi_i \chi_j 
= - \int \diff \chi_i \diff \chi_j \chi_j \chi_i = - 1$.

\paragraph{Gaussian integral.} 
There is a useful Gaussian integral formula for Grassmann variables. 
We rename our basis as $(\chi_1,\dots,\chi_q,\bar\chi_1,\dots\bar\chi_q)$.
Then for $M\in \C^{q\times q}$
\begin{align}\label{eq:susygauss}
\int \diff \bar\chi \diff \chi \eto^{-\sum_{i,j}\bar\chi_i M_{ij} \chi_j}
= \det M,
\end{align}
where $\diff \bar \chi \diff \chi 
= \prod_{j=1}^q \diff \bar\chi_j \diff \chi_j$.
Combining this with complex Gaussian integral formulas, 
we obtain the following result.

\begin{theo}[Supersymmetric integral representation]\label{theo:susyapproach}
	Let $A_1,A_2\in \C^{n\times n }$ with $\re A_1 >0$. 
	Let $\Phi = (z,\chi)^t \in \C^n \times V^n$ be a supervector and 
	$\Phi^* = (\bar z, \bar \chi) \in \C^n \times V^n$ its transpose. 
	With the notations $[\diff  \Phi^* \diff \Phi] = 
	(2\pi)^{-n} \diff \bar z \diff z \diff \bar \chi \diff \chi$ and 
	$\Phi^* A \Phi 
	= \sum_{j,k=1}^n \bar z_j (A_1)_{jk} z_k + \bar \chi_j (A_2)_{jk} \chi_k$ 
	for a block matrix 
	$A = 
	\begin{pmatrix}
	A_1 & 0 \\ 0 & A_2
	\end{pmatrix}$ 
	(a supermatrix with odd parts $0$) we can write
	\begin{align*}
	\frac{\det A_2}{\det A_1} 
	= \int [\diff \Phi^* \diff \Phi] \eto^{-\Phi^*  A \Phi}
	\end{align*}
	and
	\begin{align*}
	(A_1^{-1})_{jk} 
	=\int [\diff \Phi^* \diff \Phi] \ \bar z_k z_j\eto^{-\Phi^* \hat  A_1 \Phi},
	\end{align*}
	where 
	$\hat A_1 = 
	\begin{pmatrix}
	A_1 & 0 \\ 0 & A_1
	\end{pmatrix}$.
\end{theo}

\begin{proof}
	Combine Eq. \eqref{eq:susygauss} with the complex Gaussian integral formulas
	\begin{align*}
	\det A_1 = 
	\frac{1}{(2\pi)^n} \int \diff \bar z \diff z \eto^{-\bar z A_1 z}, 
	\quad 
	(A_1^{-1})_{jk} 
	= \frac{\det A_1}{(2\pi)^n} \int \diff \bar z  \diff z 
	\ \bar z_k z_j\eto^{-\bar z A_1 z} .
	\end{align*}
	Note that while Eq. \eqref{eq:susygauss} holds for all matrices 
	$A\in \C^{n\times n}$, we need the additional condition $\re A >0$ 
	for the complex ones to ensure that the complex integral is finite.
\end{proof}

\subsection{Grassmann algebra functions and change of variables}
In this section, we denote the body of an even element $a$  by $b(a)$ 
instead of $b_a$.
\begin{defi}\label{def:generators}
	Let $U\subset \R^p$ open. The 
	\emph{algebra of smooth $\mathcal{A}[\chi_1,\dots,\chi_q]$-valued functions} 
	on a domain $U$ is defined by 
	\begin{align*}
	\mathcal{A}_{p,q}(U)
	\coloneqq \left\{
	f = f(x,\chi)= \sum_{I\in\mathcal{P}(q)}f_I(x)\chi^I: f_I \in C^\infty (U)
	\right\}.
	\end{align*}
	We call $y_i(x,\chi)$, $ \eta_j(x,\chi)$, for ${i=1,\dots p, j = 1,\dots,q}$
	\emph{generators} of $\mathcal{A}_{p,q}(U)$
	if $p(y_i)= 0$, $p(\eta_j)=1$ and
	\begin{enumerate}[(i)]
		\item $ \{(b(y_1(x,0)),\dots, b(y_p(x,0))), x\in U\} $ is a domain in $\R^p$,
		\item we can write all $f\in\mathcal{A}_{p,q}(U)$ as $f= \sum_{I}f_I(y)\eta^I$.
	\end{enumerate}
\end{defi}
Note that $(x,\chi)$ are generators for $\mathcal{A}_{p,q}(U)$.

A change of variables is then a parity preserving transformation 
between systems of generators of $\mathcal{A}_{p,q}(U)$.
A practical change of variable formula for super integrals 
is currently only known for functions with compact support, 
i.e. functions $f\in\mathcal{A}_{p,q}(U)$ 
such that $f_I\in C_c^\infty(U)$ for all $I\in\mathcal{P}(q)$.

\begin{theo}\label{theo:changeofvariables}
	Let $U\subset\R^p$ open, $x,\chi$ and $y(x,\chi)$, $ \eta(x,\chi)$ 
	two sets of generators of $\mathcal{A}_{p,q}(U)$.
	Denote the isomorphism between the generators by 
	\begin{align*}
	\psi: (x,\chi) \mapsto(y(x,\chi),  \eta(x,\chi))
	\end{align*}
	and $V= b( \psi(U)) = \{(b(y_1(x,0)),\dots, b(y_p(x,0))), x\in U\}$. 
	Then for all $f\in  \mathcal{A}_{p,q}(V)$ with compact support, we have
	\begin{align*}%\label{eq:changeofvariables}
	\int_V \diff y \diff\eta f(y,\eta)  
	= \int_U \diff x \diff\chi f\circ \psi(x,\chi) \sdet(J\psi) ,
	\end{align*}
	where $\sdet (J\psi)$ is called the Berezinian defined by
	\begin{align*}
	J\psi = 
	\begin{pmatrix}
	\tfrac{\partial y}{\partial x} 
	& y \overleftarrow{\tfrac{\partial }{\partial \chi} }\\
	\tfrac{\partial \eta}{\partial x} 
	& \eta \overleftarrow{\tfrac{\partial }{\partial \chi}}
	\end{pmatrix}, 
	\qquad
	\sdet 
	\begin{pmatrix}
	a & \sigma \\ \rho &b
	\end{pmatrix}
	= \det (a-\sigma b^{-1} \rho) \det b^{-1}.
	\end{align*}
	Integration over even elements $x$ and $y$ means 
	integration over the body $b(x)$ and $b(y)$ 
	in the corresponding regions $U$ and $V$.
\end{theo}

\begin{proof}
	See \cite[Theorem 2.1]{Berezin1987} or \cite[Theorem 4.6.1]{Varadarajan2004}.
\end{proof}

\begin{rem}
	Applying an isomorphism $\psi$ that changes only the odd elements, 
	Theorem \ref{theo:changeofvariables} holds
	also for smooth, integrable functions 
	that are not necessarily compactly supported.
	Changing also the even elements for a non compactly supported function,
	boundary integrals can occur.
\end{rem}

\end{appendix}

%%%% bibliography
\bibliographystyle{alpha}

\end{document}